\documentclass[10pt]{article}

\usepackage[T1]{fontenc}            
\usepackage{amsmath}            
\usepackage{amssymb}                
\usepackage{amsfonts}               
\usepackage[final,pdftex]{graphicx} 
\usepackage{flafter}            
\usepackage{fullpage}               
\usepackage{titlesec}            
\usepackage{enumitem}             
\usepackage{units}                  
\usepackage{xspace}                
\usepackage{dcolumn}           
\usepackage{float}
\usepackage{cancel}
\usepackage{hyperref}

\usepackage[utf8]{inputenc}
\usepackage{authblk}

\titleformat{\section}[block]{\sffamily\bfseries\Large}{}{0pt}{}
\titleformat{\subsection}{\bfseries\sffamily}{\arabic{subsection}.}{.5em}{}
\titlespacing*{\subsection}{0pt}{2.5em}{.5em}
\title{Snowmass 2013 Young Physicists Science and Career Survey Report}
\author[a]{J. Anderson} 
\author[b]{J. Asaadi \thanks{jasaadi@fnal.gov}}
\author[a]{B. Carls} 
\author[c]{R. Cotta}
\author[d]{R. Guenette}
\author[a]{B. Kiburg}
\author[e]{A. Kobach}
\author[a]{H.~Lippincott}
\author[f]{B. Littlejohn}
\author[g]{J. Love}
\author[a.h]{B. Penning}
\author[a]{M. Soares Santos}
\author[i]{T.Strauss\thanks{thomas.strauss@lhep.unibe.ch}}
\author[d]{A.~Szelc}
\author[j]{E. Worcester}
\author[a]{F. Yu}
\affil[a]{Fermilab, Batavia, IL}
\affil[b]{Syracuse University, Syracuse, NY}
\affil[c]{University of California, Irvine, CA}
\affil[d]{Yale University, New Haven, CT}
\affil[e]{Northwestern University, Evanston, IL }
\affil[f]{University of Cincinnati, Cincinnati, OH }
\affil[g]{Argonne National Laboratory, Argonne, IL}
\affil[h]{University of Chicago, Chicago, IL}
\affil[i]{University of Bern, Bern, Switzerland}
\affil[j]{Brookhaven National Lab, Upton, NY}

\begin{document}
  \maketitle
\thispagestyle{empty}
\begin{abstract} From April to July 2013 the Snowmass Young Physicists (SYP) administered an online survey collecting the opinions and concerns of the High Energy Physics (HEP) community. The aim of this survey is to provide input into the long term planning meeting known as the Community Summer Study (CSS), or Snowmass on the Mississippi. In total, 1112 respondents took part in the survey including 74 people who had received their training within HEP and have since left for non-academic jobs. This paper presents a summary of the survey results including demographic, career outlook, planned experiments, and non-academic career path information collected.
\end{abstract}
\newpage
\pagenumbering{roman}
\tableofcontents{}
\newpage
 \pagenumbering{arabic}
 
\section{Introduction}

The Snowmass Young Physicists (SYP) \cite{YPM2013:Website} was first organized at the Community Planning Meeting held at Fermilab in October of 2012. This group was formed to provide a conduit for young (nontenured) particle physicists to participate in the forthcoming Community Summer Study  (known as Snowmass on the Mississippi) which took place in the summer of 2013 in Minneapolis, MN.

The primary charge taken up by the SYP was to facilitate and encourage young people to get involved  with physics studies and meetings in preparation for Snowmass on the Mississippi. The SYP generated an online platform (http://snowmassyoung.hep.net) as well as an offline network for advertising tasks that need to be done and connecting interested SYP members with the relevant frontiers. This was accomplished by having our members attend many of the ``Pre-Snowmass'' meetings across the frontiers. While at these meetings, dedicated parallel sessions were arranged with remote participation to provide a platform for discussion and input. Invited speakers from the Fermilab directorate, Department of Energy (DOE), National Science Foundation (NSF), as well as senior scientists and faculty came to speak with a broad section of the young community. In addition to these meetings, over a dozen SYP town hall meetings were held prior to Snowmass during which plans and concerns were raised.

The second charge is related to a ``deliverable'' to the Community Summer Study. The SYP gathered information about demographics, career outlook, physics outlook, planned experiments, and general concerns of young physicists in the form of an online survey \cite{survey:2013}. This information is intended to provide a voice to the next generation of leaders in High Energy Physics (HEP) and serve as a basis for discussion with senior physicists, politicians, and funding agencies about the current and future state the field. Additionally the opinions expressed here represent many people who were unable to attend Snowmass on the Mississippi personally. The results of this survey consititute the majority of this paper.

The next charge of the SYP is to become a long term asset to young physicists. This is done through multiple channels such as providing information and resources to people in high energy physics when they are making career decisions. This includes, but is not limited to, information about current and planned scientific experiments and collaborations, having cross-frontier talks and seminars, and providing information and resources for those who decide to take the many skills learned in physics out into the general work force. Furthermore, the SYP will make efforts to engage members of Congress from the many districts in which the SYP members live and help lobby for the interests of young people in HEP. 

Moreover, the SYP aims to provide a chance for young physicists to network and meet each other as well as become known outside of each of our particular subfields.

Beyond the above listed tasks, SYP has made contact with individuals who have recently left HEP for non-academic career paths.  Through gathering information about their experiences we can illustrate the broader positive impact members of HEP have on the wider U.S. after leaving the field. Moreover this information can serve as an example for career opportunities of current would-be physicists in the future. In addition to the online survey that was taken leading up to Snowmass, a follow-up document is availabe for people who know the contact information for those who have left HEP (\url{http://snowmassyoung.hep.net/OA_contacts.pdf}). 

The mechanism of young scientists organizing to provide input into the Snowmass process is not a new phenomenon, as a group organized during Snowmass 2001 and became known as the Young Particle Physicists (YPP) \cite{YPP:2001}. The YPP also administered a survey in order to collect the opinions and concerns of many in HEP \cite{Fleming:2001zk}. This work became the basis for the SYP in 2013 who were in fact helped a great deal by these original scientists, many of whom have moved into leadership roles across HEP.

This paper is dedicated to showing a summary of the results of the SYP 2013 survey. The data on which these results are based is now publicly available (\url{http://snowmassyoung.hep.net/SYP2013_PublicData.tar.gz}). Section \ref{sec:Method} will outline the methodology of the survey. Section \ref{sec:Demo} will present the results from the demographic information collected in the survey. In Section \ref{sec:careerOutlook} we present results pertaining to the opinions of peoples career outlook in HEP.  More generally, subsection \ref{sec:spires} helps frame this discussion in terms of available and filled jobs as reported by Spires data \cite{Spires:Data} \cite{Spires:RawData}. Section \ref{sec:physicsOutlook} presents results pertaining to opinions on the physics prospects in HEP including information about which of the planned experiments appear most exciting to the community. In Section \ref{sec:NonAcademic} we present some of the opinions gathered by those who received their training within HEP but have chosen to pursue careers outside of academia. Finally in Section \ref{sec:surveyquestions} we present all the survey questions and include some of the direct responses we received from the many participants.

\section{Survey Methodology} \label{sec:Method}

The goal of the survey administered by the Snowmass Young Physicists was to collect a broad range of opinions that reflect the physics interests, the career outlook, and the general mood of the field leading up to Snowmass 2013. The survey was administered using Google Forms \cite{Google:Forms} and available from April 1st to July 15th 2013 (\url{http://tinyurl.com/snowmassyoung}).

This survey contains four distinct sections:

\begin{enumerate}

\item[I] \textbf{Demographic Information:} This includes general information such as the gender, marital status, and citizenship but also information that is specific to the demographics of HEP. This information includes current frontier of work, current position within academia (e.g. graduate student, post-doctoral researcher, tenured faculty, etc...), and number of years in your current position. Similar information for those on a non-academic career path was also gathered.

\item[II] \textbf{Career Outlook:} This section asked the respondent questions pertaining to their feelings and outlook on their current career. Questions included what type of jobs are most attractive to young scientists, what factors impact their decisions to pursue an academic career, as well as what various external factors (family, availability of jobs, job location, etc...) that impact their career decisions.

\item[III] \textbf{Physics Outlook:} These questions focused on the science that is being planned during Snowmass 2013. For example, respondents were asked: what their outlook for funding of HEP in the future is, which frontier will have the greatest impact in the next 10 years on HEP, and to indicate which of the planned experiments (given from a non-exhaustive list) did the respondents have the highest priority for.

\item[IV] \textbf{Non-academic Career Paths:} Finally, this section of the survey was dedicated to gathering information from those who had received their training within HEP but have since left to pursue a career outside of academia. In this section we asked questions about how their time in HEP prepared them for their current career. Additionally we gathered information to see how their professional lives compare to those working within HEP.
\end{enumerate}

\begin{figure}[H]
  \begin{center}
    \includegraphics[scale=0.50]{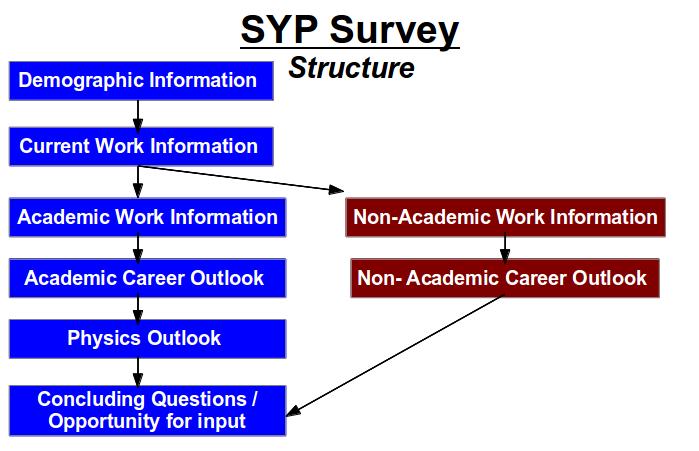}
    \caption{Flow chart for the SYP survey. All respondents saw the same general demographic, work, and concluding questions. However, those who have chosen a non-academic career path were exposed to a different set of questions and not asked questions about current physics. \label{fig:SurveyFlowChart}}
 \end{center}
\end{figure}

The survey was structured such that respondents would only see certain questions based on information about their current employment. Figure \ref{fig:SurveyFlowChart} shows a summary of the flow chart for the survey. This allows us to look for similarities across the different career paths without exposing respondents to questions that had little relevance to them. In hindsight, it would have been helpful to do the same for tenured faculty and young people, since many questions asked about preparing for careers outside of HEP, and had little relevance for those more senior members of our field. For the survey that is to be administered post-Snowmass, \url{http://tinyurl.com/post-snowmass}, this issue has been resolved. 

The parsing of responses to the survey allowed us to identify correlations across different frontiers and different career stages. While many of the questions from the Career Outlook section focused on concerns most relevant to young scientists, most of the questions and answers included options that could be answered by all who were taking the survey.

A small amount of data clean up was necessary before the sample could be analyzed. This was mostly to format answers to questions in such a way that they could be analyzed using C++ scripts. An example of one such clean up came from a question asking users to input the average number of hours worked in a week. Many users inputted ``$\sim$ 45''. This tilde was simply removed from the data before being outputted to C++ code.

Finally results of the survey were outputted to a series of ROOT files which can be read by a simple analysis script. A complete survey response is seen as an ``event'' and this allows the end user to write a simple analysis script to apply selection, make histograms, and output the results. A complete package including the raw data, translated ROOT files, and an example analysis script can be found here:

\begin{center}
\url{http://snowmassyoung.hep.net/SYP2013_PublicData.tar.gz}
\end{center}

The plots shown in this paper have negligible uncertainties and a full set of plots with uncertainties can be found here:

\begin{center}
\url{http://snowmassyoung.hep.net/plots.html}
\end{center}

\section{Demographic}\label{sec:Demo}

Demographic information composed an important part of the survey and included $\sim$ 10-12 questions (depending on if you were on the academic or non-academic path of the survey) including information about gender, marital status, number of children, and current employment.

A total of 1112 responses were collected for the SYP 2013 survey, of which 956 fit the ``young'' definition, e.g. non-tenured inside academia, and 74 respondents coming from non-academic career paths. Comparing this to the 2001 survey which collected 1508 responses, of which 857 fit the ``young'' definition, we can see that we reached fewer total respondents, but a larger fraction of respondents were young people.

Below is a summary of some general demographic information collected and, where relevant, comparisons to U.S. 2010 census data \cite{census:2010} are made.

\begin{enumerate}

\item \textbf{What is your gender?}

Male: \textbf{79$\%$} (U.S. 2010 Census Data: 49$\%$)

Female: \textbf{21$\%$} (U.S. 2010 Census Data: 51$\%$)

\begin{figure}[H]
\begin{center}
\includegraphics [scale=0.30]{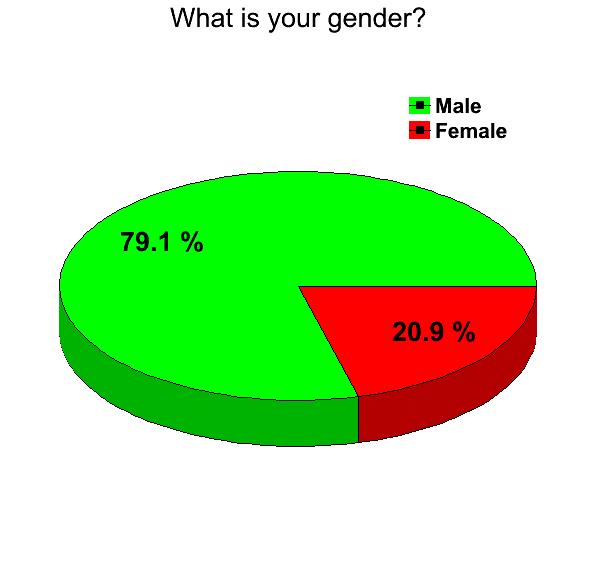}
\caption{Response to demographic question about gender.}\label{fig:gender}
\end{center}
\end{figure}

\item \textbf{What is your current marital status?}

Married: \textbf{38$\%$} (U.S. 2010 Census Data for population $>$ 18 years old: 51$\%$)

Un-Married: \textbf{62$\%$} (U.S. 2010 Census Data for population $>$ 18 years old: 49$\%$)

\begin{figure}[H]
\begin{center}
\includegraphics [scale=0.30]{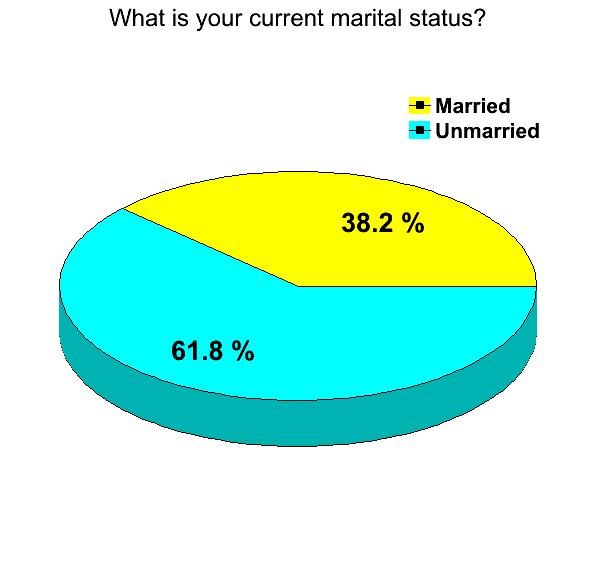}
\caption{Response to demographic question current marital status.}\label{fig:marriage}
\end{center}
\end{figure}

\item \textbf{How many children do you have?}

No children: \textbf{79$\%$} (U.S. 2010 Census Data for population $>$ 18 years old: 52$\%$)

$\geq$ 1 child \textbf{21$\%$} (U.S. 2010 Census Data for population $>$ 18 years old: 48$\%$)

\begin{figure}[H]
\begin{center}
\includegraphics [scale=0.25]{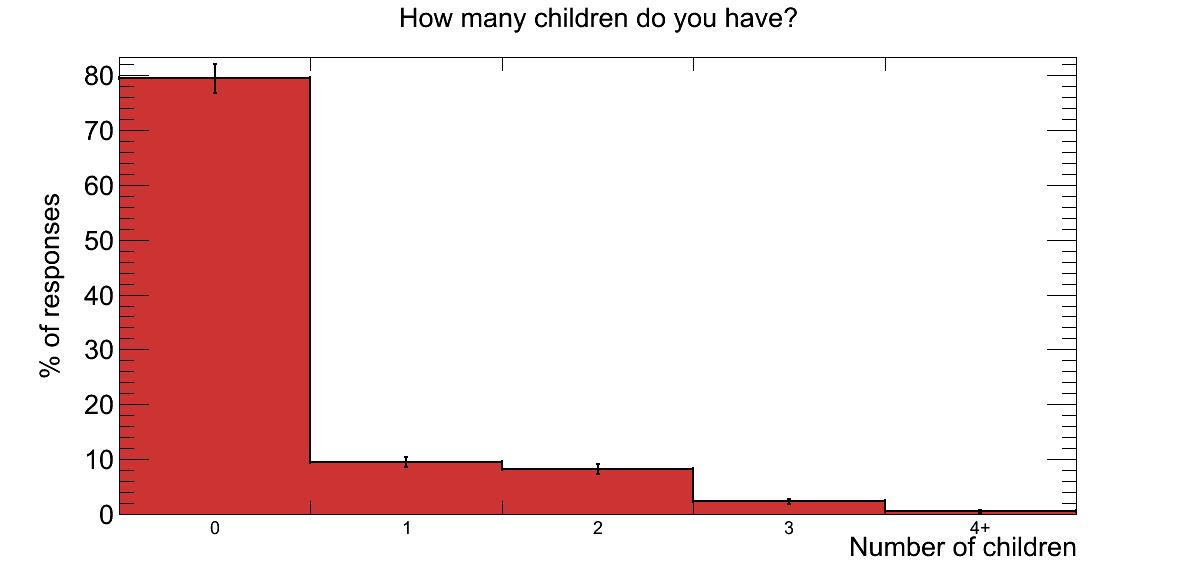}
\caption{Response to demographic question about number of children.}\label{fig:kids}
\end{center}
\end{figure}

\item \textbf{What is your household salary (USD/year)?}

\textbf{Making $<$ 75,000 USD/year:} 69$\%$ (U.S. 2010 Census Data for population $>$ 18 years old: 80$\%$) 

\textbf{Making $>$ 75,000 USD/year:} 31$\%$ (U.S. 2010 Census Data for population $>$ 18 years old: 20$\%$) 

\begin{figure}[H]
\begin{center}
\includegraphics [scale=0.25]{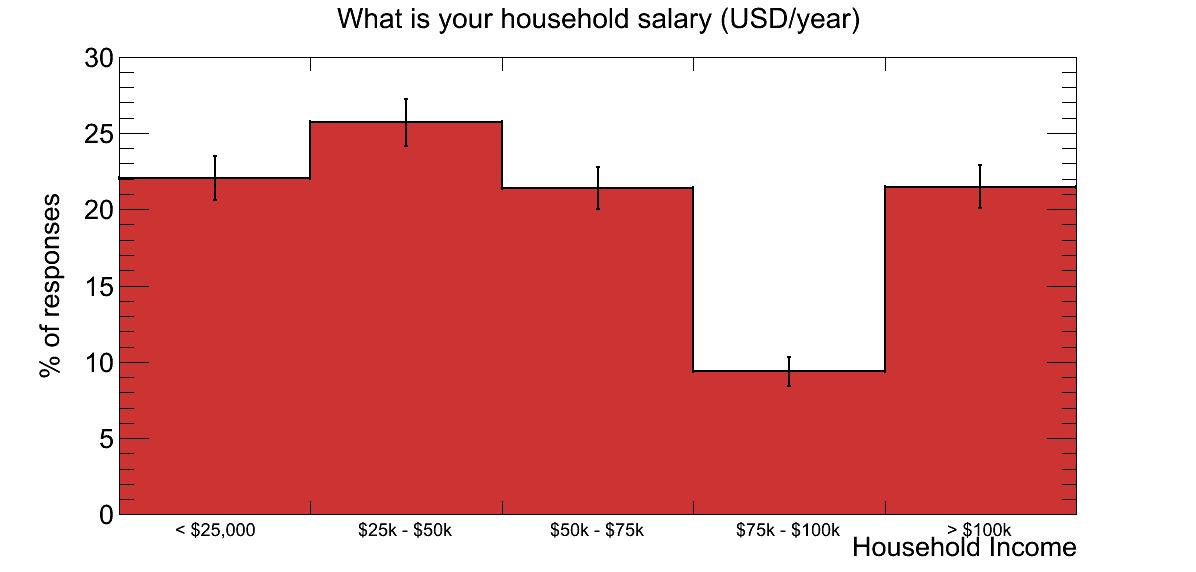}
\caption{Response to demographic question about household income.}\label{fig:salary}
\end{center}
\end{figure}

\end{enumerate}

\begin{figure}[H]
\begin{minipage}[c]{.475\linewidth}
\begin{center}
\includegraphics [height=6cm,width=9cm]{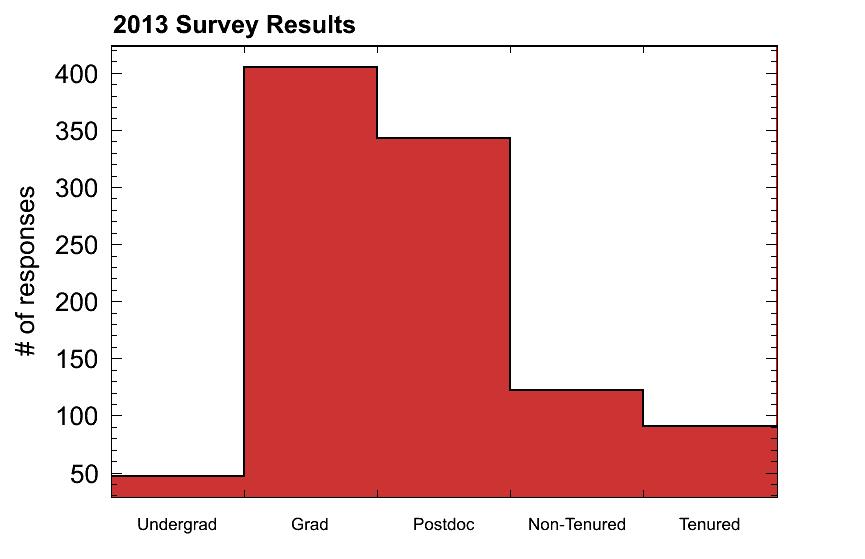}
\caption{Current position of the survey participants 2013.}\label{fig:currentpostion2013}
\end{center}
\end{minipage}
\hspace{.05\linewidth}
\begin{minipage}[c]{.475\linewidth}
\begin{center}
\includegraphics[scale=0.31]{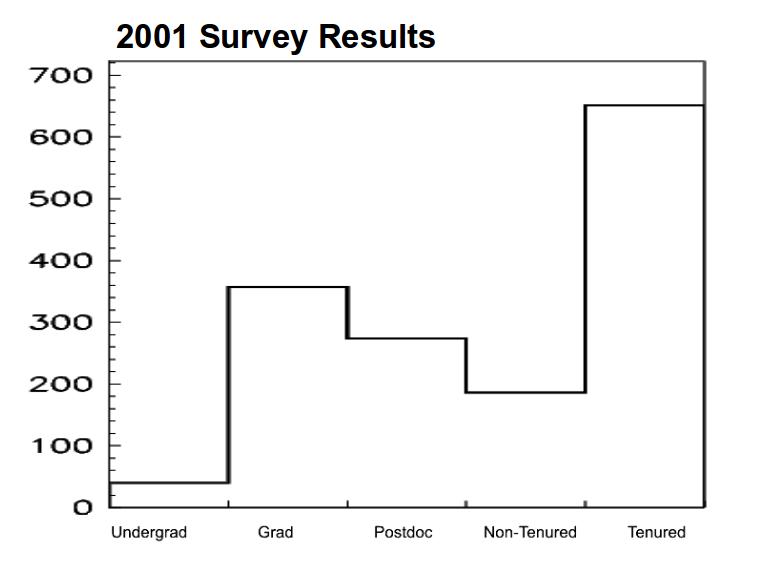}
\caption{Current position of the survey participants 2001.}\label{fig:currentpostion2001}
\label{currentpos_2001}
\end{center}
\end{minipage}
\end{figure}

Exploring the demographic information further, Figure \ref{fig:currentpostion2013} shows the breakdown of the respondents whom are currently working in academia by their current position. Figure \ref{fig:currentpostion2001} shows this same break down of the HEP respondents by their academic positions in the 2001 survey.

\begin{figure}[H]
\begin{center}
\includegraphics [scale=0.30]{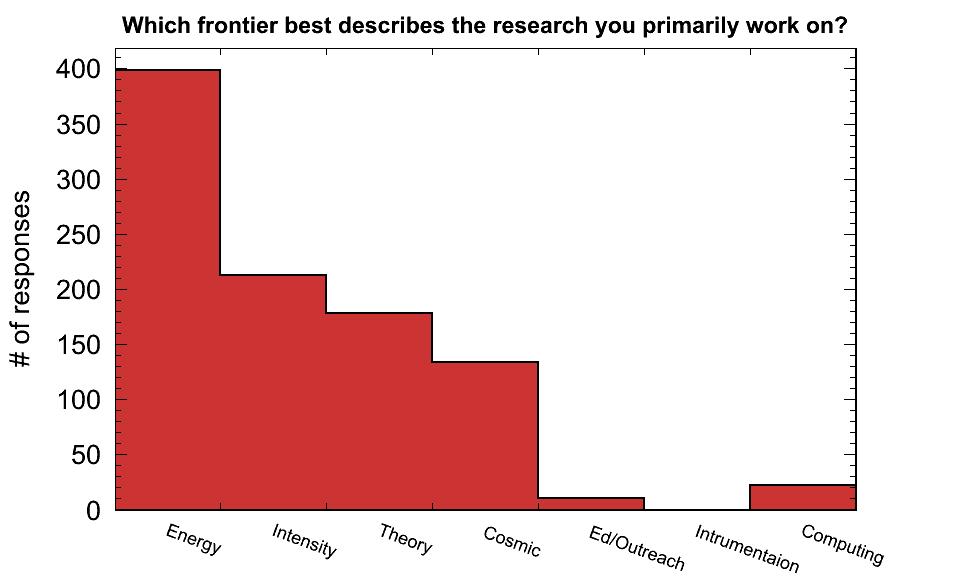}
\caption{Current research frontier of the survey participants as defined by the Snowmass process.}\label{fig:frontier}
\end{center}
\end{figure}

\begin{figure}[H]
\begin{center}
\includegraphics [scale=0.35]{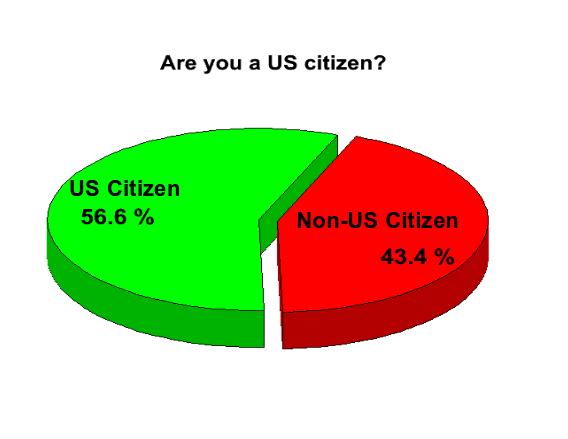}
\caption{Citizenship of the survey respondents.}\label{fig:citizen}
\end{center}
\end{figure}

The Snowmass process has identified 7 "frontiers" in HEP: Energy, Intensity, Theory, Cosmic, Education/Outreach, Instrumentation, and Computing. Figure \ref{fig:frontier} shows the number of respondents who primarily work in each of those frontiers. A large number of participants come from the Energy Frontier, with the Intensity, Theory, and Cosmic Frontiers making up the majority of the rest. Figure \ref{fig:citizen} shows breakdown of survey participants by their current citizenship, with over half coming from the United States.


\begin{figure}[H]
\begin{center}
\includegraphics [scale=0.30]{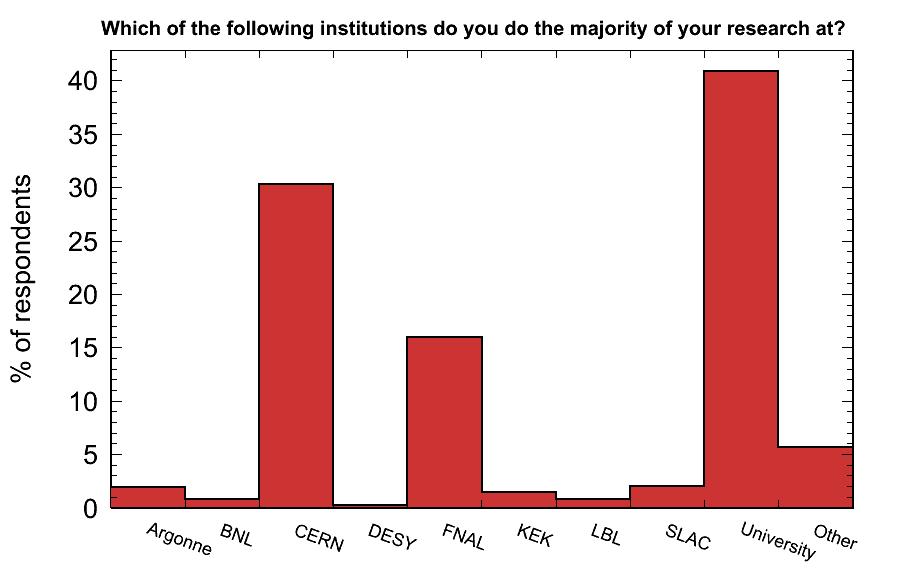}
\caption{Institution respondents do the majority of their research at.}\label{fig:whereResearch}
\label{exitment_2013}
\end{center}
\end{figure}

Figure \ref{fig:whereResearch} shows where participants do the majority of their research. The largest number of respondents are based at universities, followed by CERN and FNAL. This is a shift from the responses found in the 2001 survey which had the majority of their responses coming from users at Fermilab \cite{Fleming:2001zk}.

\begin{figure}[H]
\begin{minipage}[c]{.475\linewidth}
\begin{center}
\includegraphics [scale=0.35]{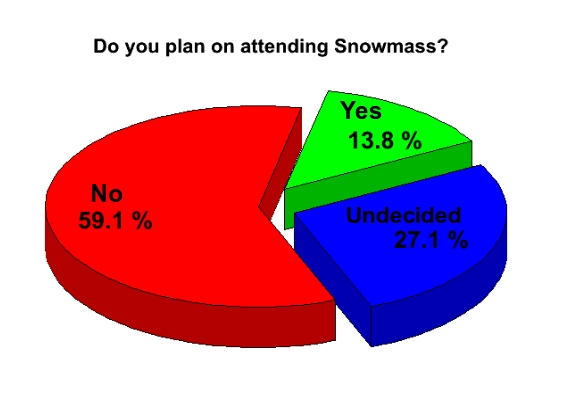}
\caption{Planned attendance at Snowmass 2013.} \label{fig:attend}
\end{center}
\end{minipage}
\hspace{.05\linewidth}
\begin{minipage}[c]{.475\linewidth}
\begin{center}
\includegraphics [scale=0.30]{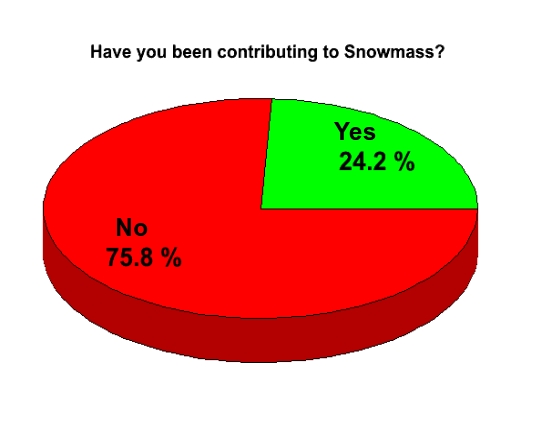}
\caption{Fraction of respondents contributing to the Snowmass process.}
\label{fig:contribute}
\end{center}
\end{minipage}
\end{figure}

Finally, Figure \ref{fig:attend} shows the breakdown of the survey participants that plan on attending the Snowmass meeting. Almost 60$\%$ of those responding did not plan on attending the Snowmass meeting with another 27$\%$ undecided. Figure \ref{fig:contribute} shows that fewer than one-fourth of survey respondents were contributing to Snowmass prior to taking the survey.

The demographic information tells us that the survey reached mostly young scientists in HEP, many of whom do their research at universities and who had not been participating in the Snowmass process prior to taking this survey. Moreover, many of those taking the survey will not be attending the Snowmass meeting. This makes the data presented here particularly relevant as it expresses the opinions and concerns of many who otherwise have have not been involved in the Snowmass process.

\section{Career Outlook} \label{sec:careerOutlook}

Career outlook questions were generally focused on the opinions and outlook of those pursuing a career within HEP. Information about the outlook for future funding and its impact on career decisions as well as questions about the dominant factors that impact decisions to pursue an academic career were covered.

Some general trends we observed about the career outlook of those in HEP responding to our survey:

\begin{enumerate}

\item \textbf{On a scale of 1 to 10 (1 = Funding will stop, 10 = Funding will thrive) how do you feel about the funding within your frontier within the next decade?}

Nearly 60$\%$ of the respondents believed that funding was more likely to decline (giving an answer less than or equal to 5) in the future.

\begin{figure}[H]
\begin{center}
\includegraphics [scale=0.25]{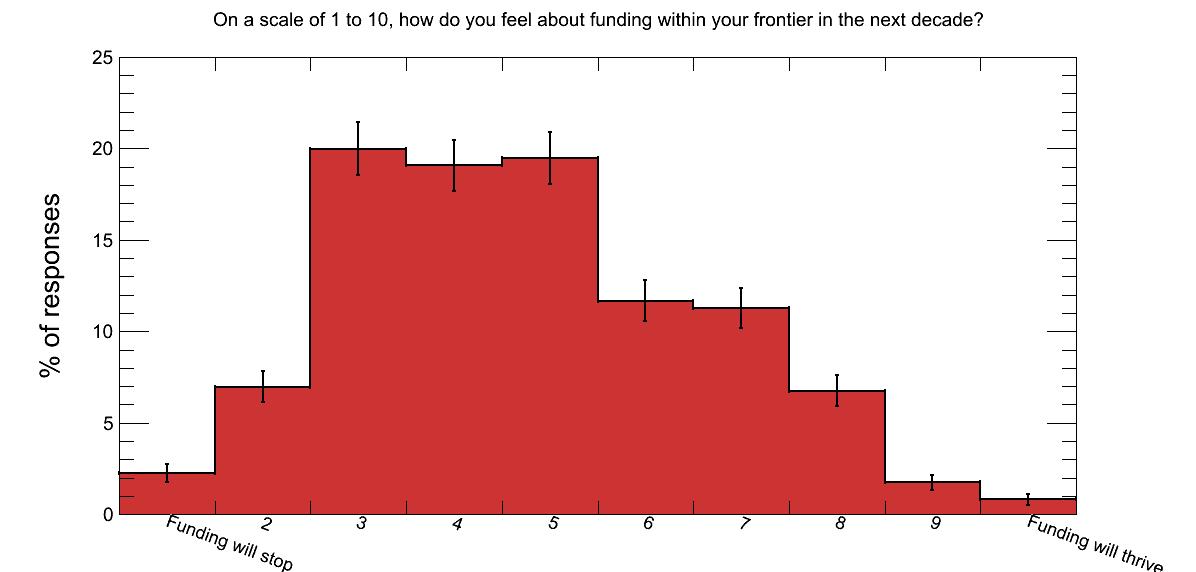}
\caption{Response to question about the future of funding within the respondents frontier in the next decade.}\label{fig:funding}
\end{center}
\end{figure}

Figure \ref{fig:fundingyoungold} shows the breakdown of this question by the classification of young and senior. This demonstrates that the young respondents are slightly more pessimistic about the future funding than their senior counterparts.

\begin{figure}[H]
\begin{center}
\includegraphics [scale=0.25]{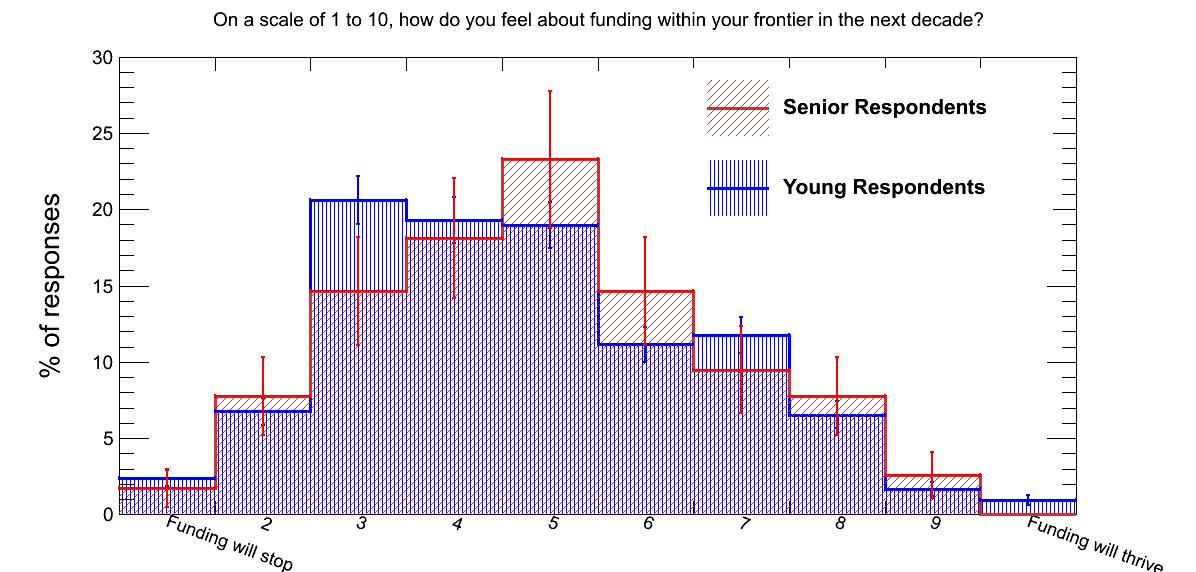}
\caption{Response to question about the future of funding within the respondent's frontier in the next decade broken down by young and senior responses.}\label{fig:fundingyoungold}
\end{center}
\end{figure}

\item \textbf{Do you intend to pursue a permanent career in academia?}

Despite this knowledge about a bleak funding profile being well known amongst the young scientists, $\sim$62$\%$ of the respondents still intend to pursue a career in academia.

\begin{figure}[H]
\begin{center}
\includegraphics [scale=0.30]{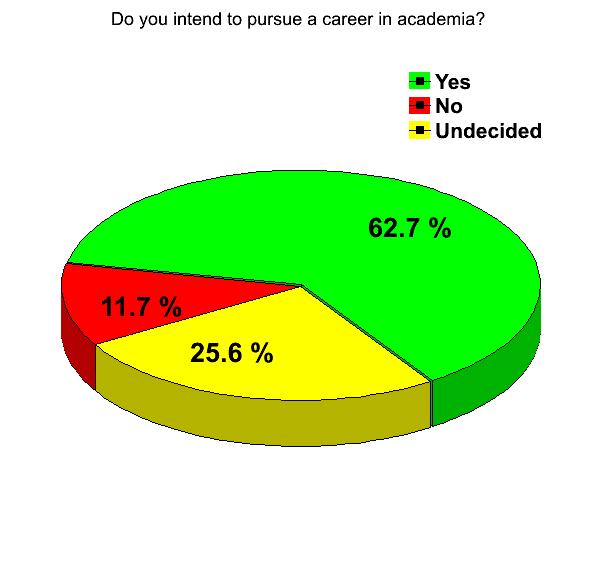}
\caption{Response to question about the future of funding within the respondent's frontier in the next decade broken down by young and senior responses.}\label{fig:pursueacademia}
\end{center}
\end{figure}

\item \textbf{Which of the following career related concerns do you find the most important to you and your future in high energy physics?}

The two most important career related concerns were the availability of university based jobs followed by the availability of laboratory based jobs.

\begin{figure}[htp]
\begin{minipage}[c]{.475\linewidth}
\begin{center}
\includegraphics [scale=0.20]{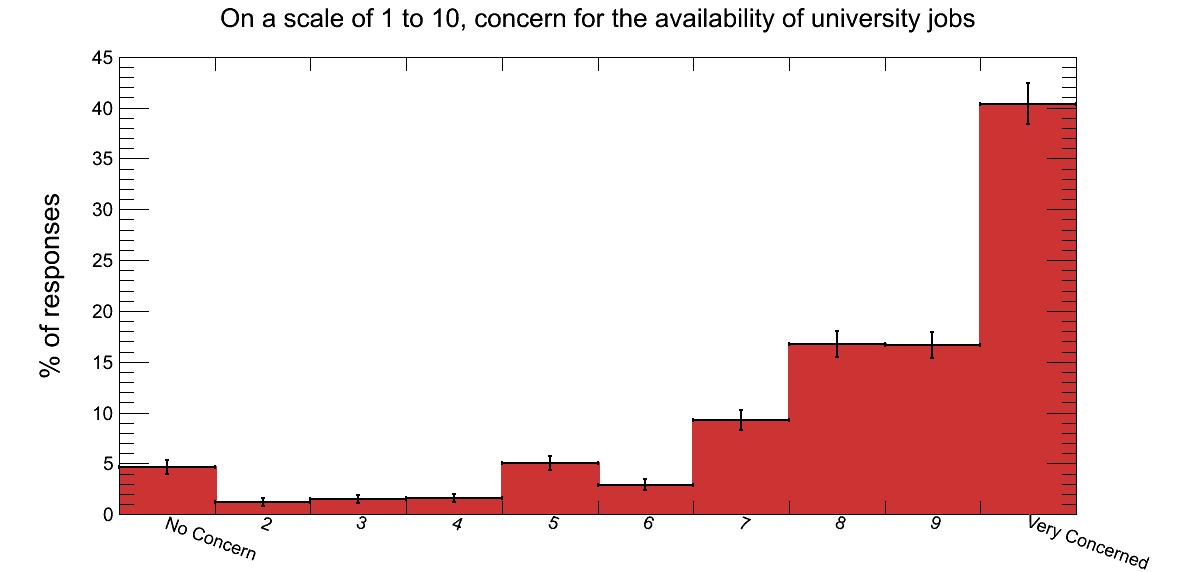}
\caption{Response to question about the impact of the availability of university jobs to the career and future of the respondent in HEP.} \label{fig:UniversityJob}
\end{center}
\end{minipage}
\hspace{.05\linewidth}
\begin{minipage}[c]{.475\linewidth}
\begin{center}
\includegraphics [scale=0.20]{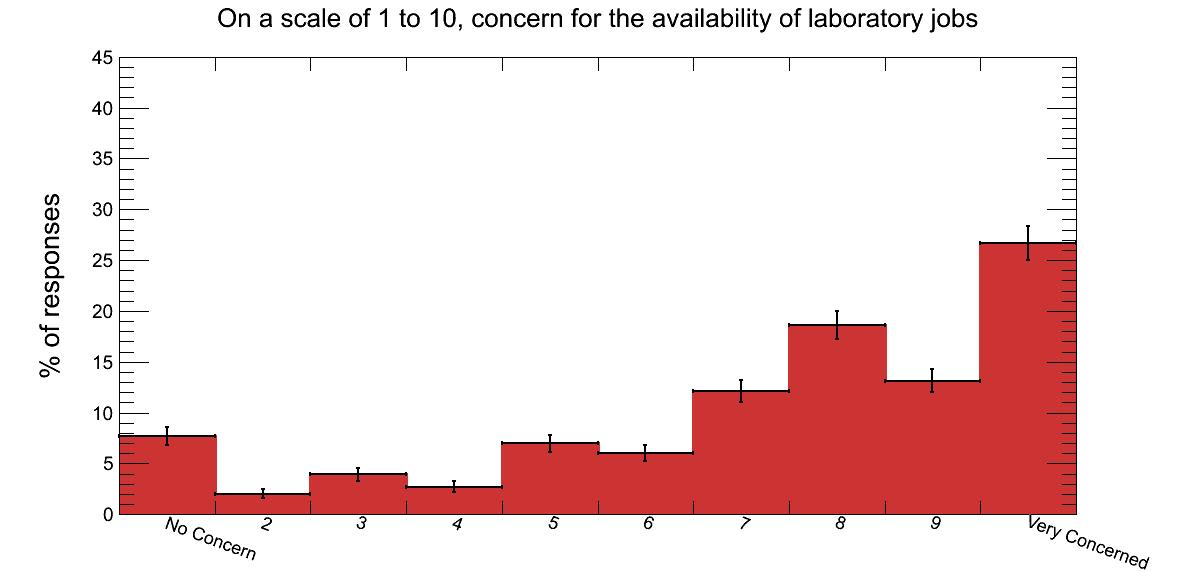}
\caption{Response to question about the impact of the availability of laboratory jobs to the career and future of the respondent in HEP.} \label{fig:LabJob}
\label{fig:contribute}
\end{center}
\end{minipage}
\hspace{.05\linewidth}
\begin{minipage}[c]{.475\linewidth}
\begin{center}
\includegraphics [scale=0.20]{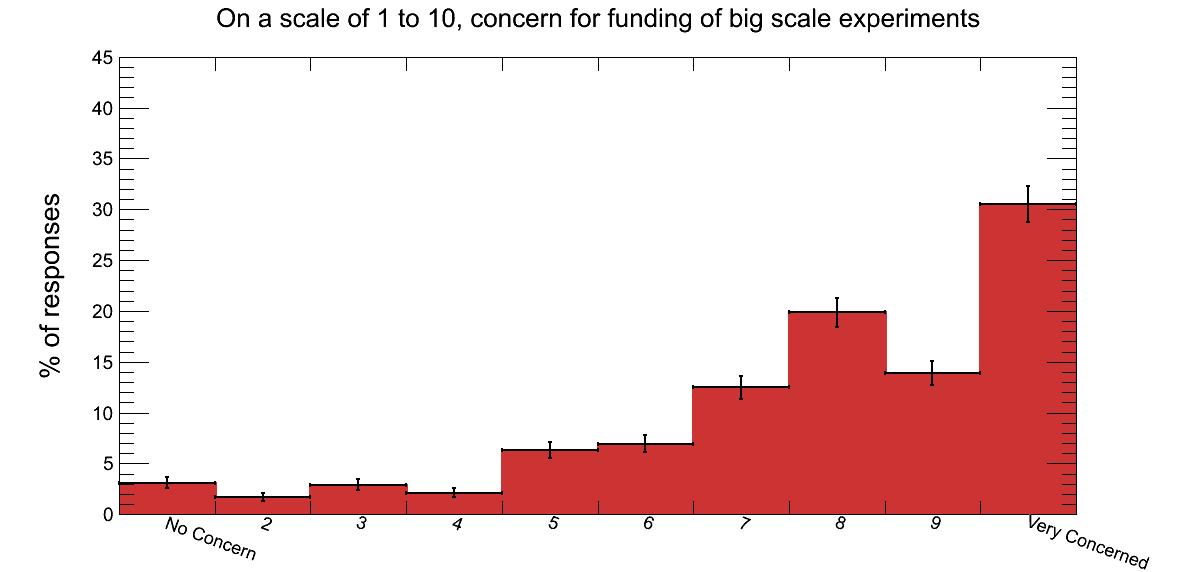}
\caption{Response to question about the impact of funding for big scale experiments to the career and future of the respondent in HEP.} \label{fig:Bigscale}
\label{fig:contribute}
\end{center}
\end{minipage}
\hspace{.05\linewidth}
\begin{minipage}[c]{.475\linewidth}
\begin{center}
\includegraphics [scale=0.20]{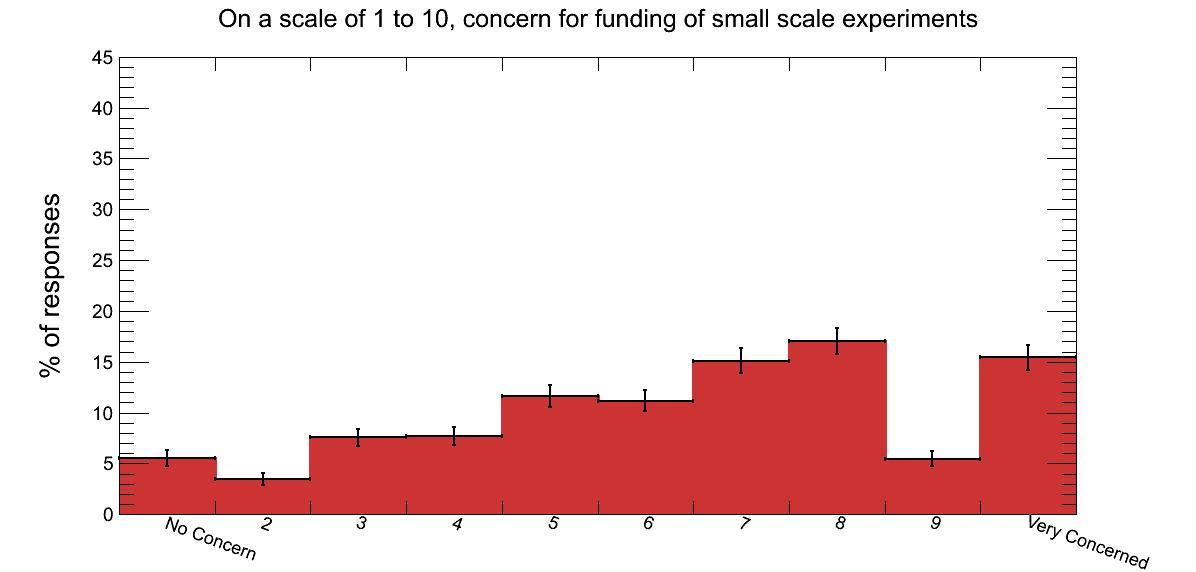}
\caption{Response to question about the impact of funding for small scale experiments to the career and future of the respondent in HEP.} \label{fig:Smallscale}
\label{fig:contribute}
\end{center}
\end{minipage}
\end{figure}

\end{enumerate}

\begin{figure}[htp]
\begin{center}
\includegraphics [scale=0.30]{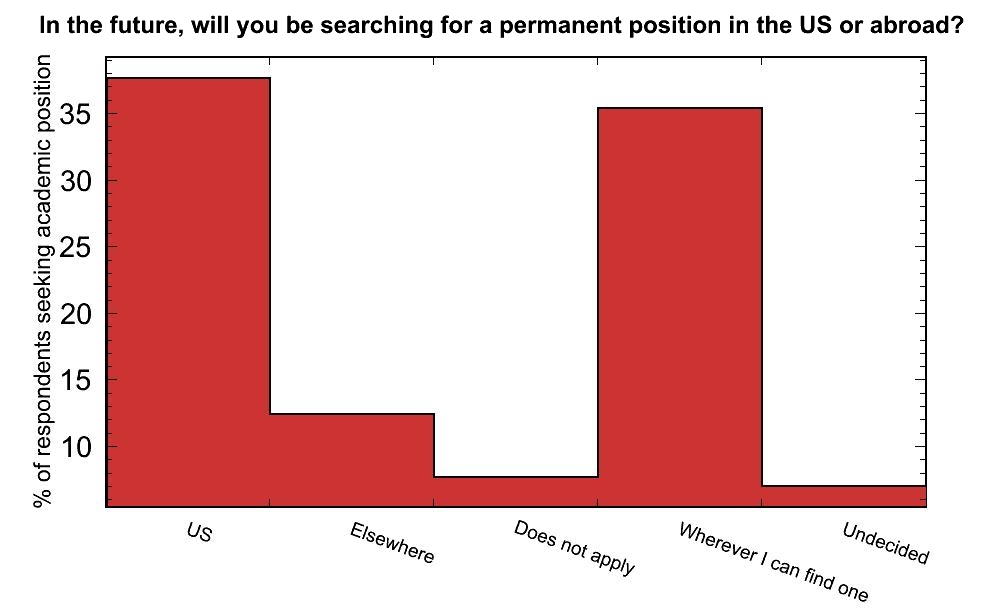}
\caption{Responses from those who intend to seek an academic career to the question of where the survey participant will be searching for a permanent position.} \label{fig:JobSearchAll}
\end{center}
\end{figure}

In addition to the general trends from the questions listed above, Figure \ref{fig:JobSearchAll} shows responses to the question of where young scientists who intend to seek an academic position will search for a permanent position. This shows that the majority intend to search for a permanent position within the United States with the next largest group willing to search wherever they can find at job. 

\begin{figure}[htp]
\begin{center}
\includegraphics [scale=0.30]{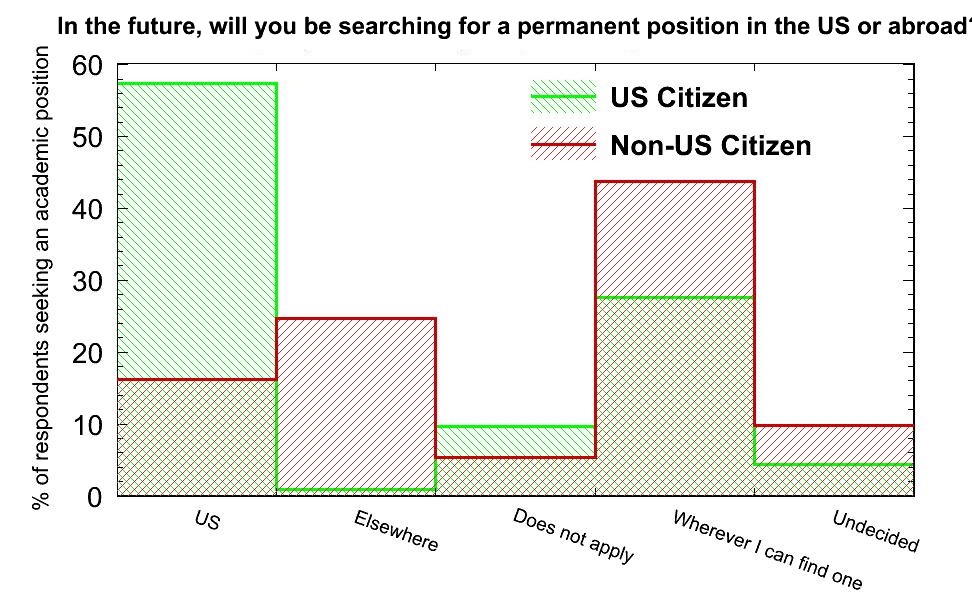}
\caption{Responses from those who intend to seek an academic career to the question of where the survey participant will be searching for a permanent position broken down by those who are U.S. citizens and those who are non-U.S. citizens.}
\label{fig:JobSearchUSNonUS}
\end{center}
\end{figure}

Figure \ref{fig:JobSearchUSNonUS} shows the breakdown of this question based on whether the person in a U.S. or non-U.S. citizen.  Clearly,  U.S. citizens prefer to search for a job within the U.S. while non-U.S. citizens are more apt to take a job wherever they can find one. One possible interpretation to this breakdown could be that in order to attract the best of the non-U.S. members of the HEP community, job availability is already a driving force. 

\begin{figure}[htp]
\begin{center}
\includegraphics [scale=0.30]{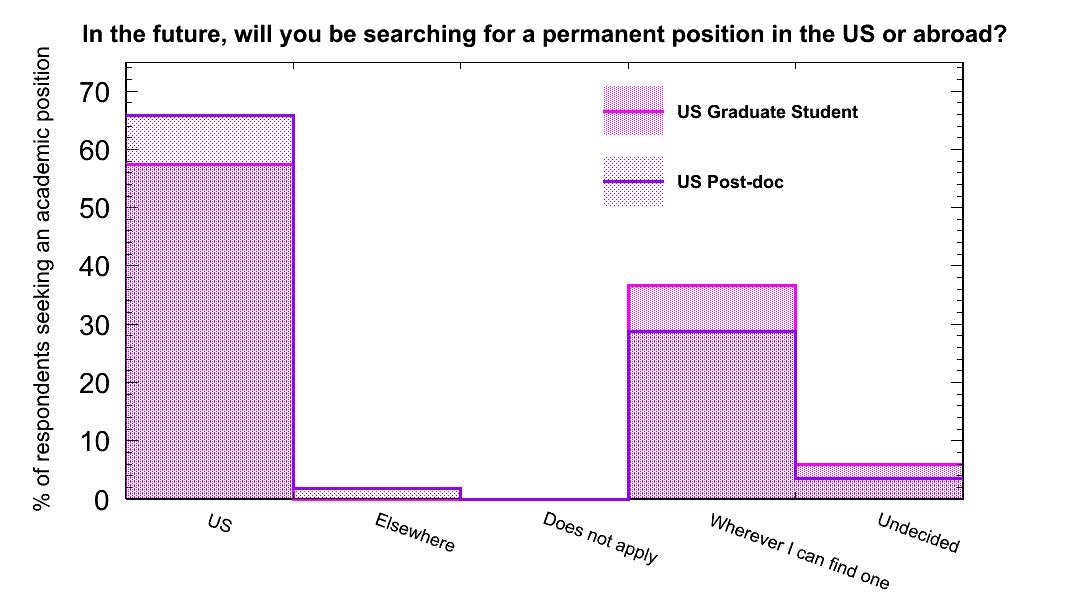}
\caption{Responses from those who intend to seek an academic career to the question of where the survey participant will be searching for a permanent position broken down by those who are currently U.S. graduate students and U.S. Postdoc.}
\label{fig:JobSearchGradPostdoc}
\end{center}
\end{figure}

\begin{figure}[htp]
\begin{center}
\includegraphics [scale=0.30]{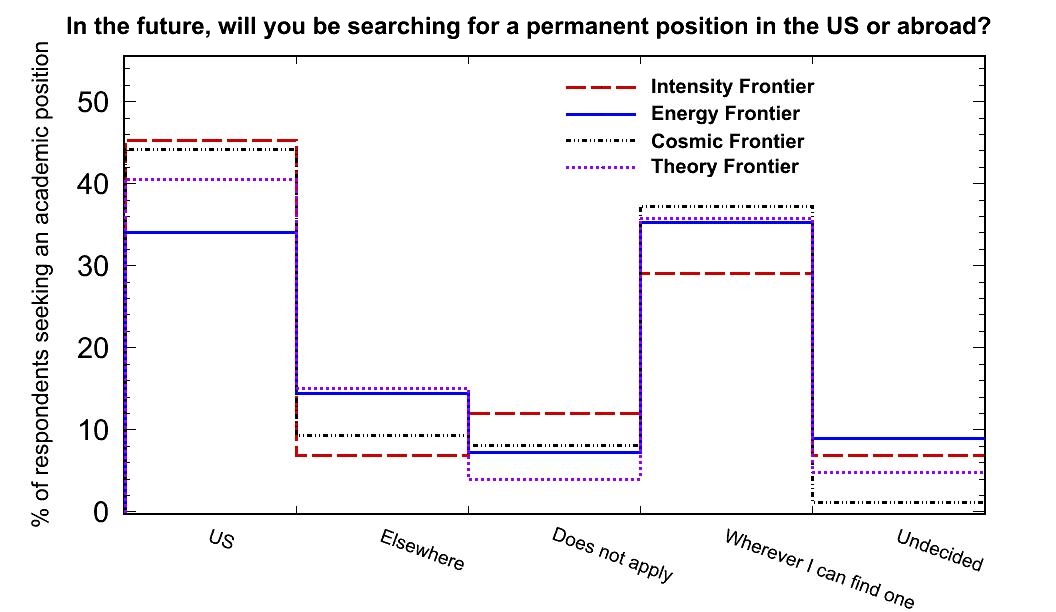}
\caption{Responses from those who intend to seek an academic career to the question of where the survey participant will be searching for a permanent position broken down by various frontiers.}
\label{fig:JobSearchFrontier}
\end{center}
\end{figure}

Figure \ref{fig:JobSearchGradPostdoc} suggests that the younger generation of U.S. scientists are more willing to take a job wherever they can find one compared to the Post-doctoral respondents. Figure \ref{fig:JobSearchFrontier} shows the breakdown to this question based on frontier, indicating that people working in the Intensity Frontier are more apt to search for a job in the U.S. while those from the Theory, Cosmic, and Energy Frontiers are more open to accepting a job wherever they can find one. These trends again suggest that a major driving force for where the young and bright physicists will go is where the jobs are available.

\begin{figure}[htp]
\begin{center}
\includegraphics [scale=0.30]{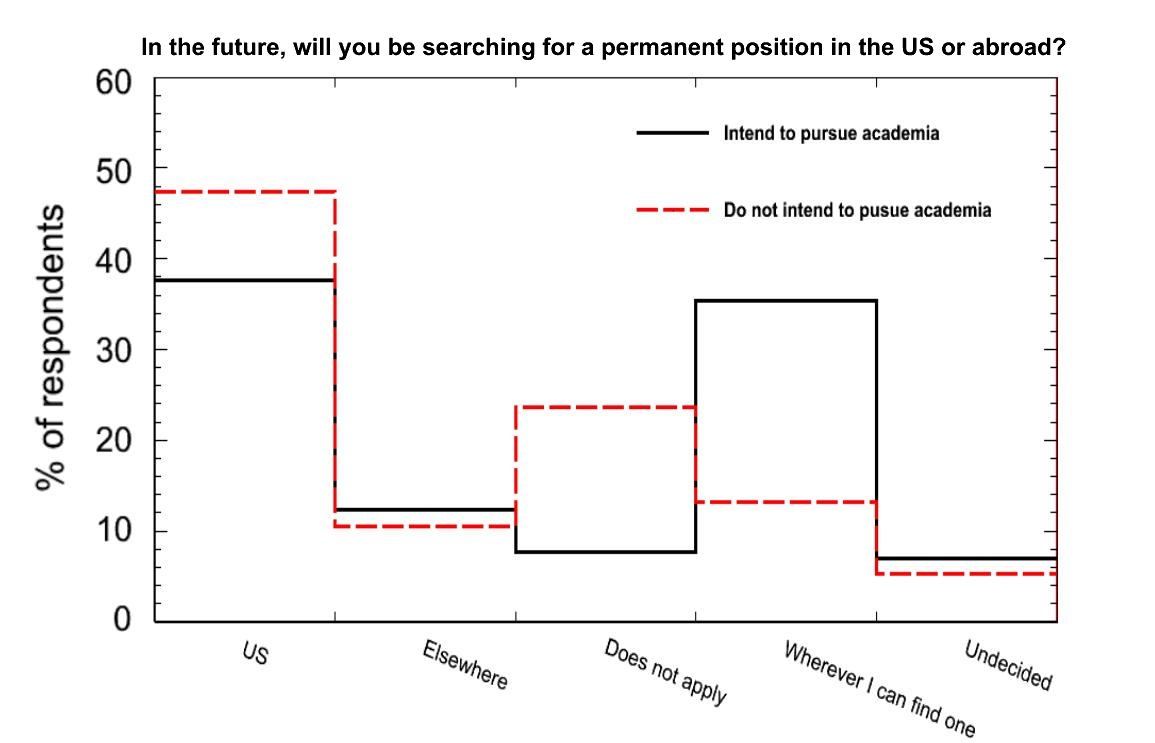}
\caption{Responses to the question of where will the survey participant be searching for a permanent position broken down by those who intend to search for a job in academia.} \label{fig:JobSearchAcademia}
\end{center}
\hspace{.05\linewidth}
\end{figure}

A further confirmation of this general trend can be seen in Figure \ref{fig:JobSearchAcademia}. This figure shows where our respondents intend to search for a job broken down by whether they intend to pursue an academic job. The results indicate that those who wish to pursue a career in academia are more willing to go wherever the jobs are, with less regard for where those jobs are. 

To follow-up on this observed trend we show the responses from those who intend to seek an academic career to the question if they will be more inclined to search for a job outside the U.S. if the next major experiment in their frontier is built outside the U.S.

\begin{figure}[htp]
\begin{center}
\includegraphics [scale=0.30]{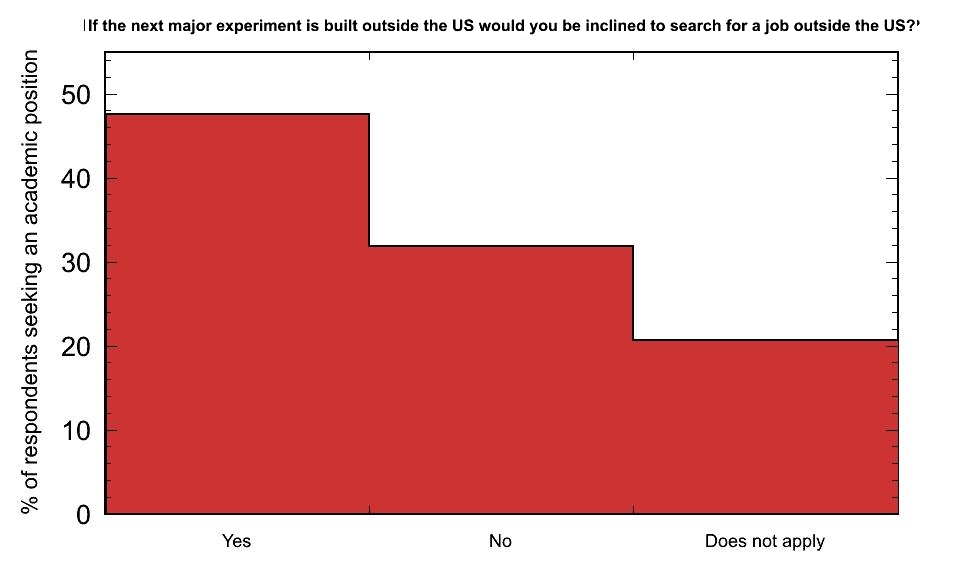}
\caption{Response from those who intend to seek an academic career to the question if they would be inclined to search for a job outside the U.S. if the next major experiment in their field was built outside the U.S.} \label{fig:LeavingUSAll}
\end{center}
\end{figure}

\begin{figure}[htp]
\begin{center}
\includegraphics [scale=0.30]{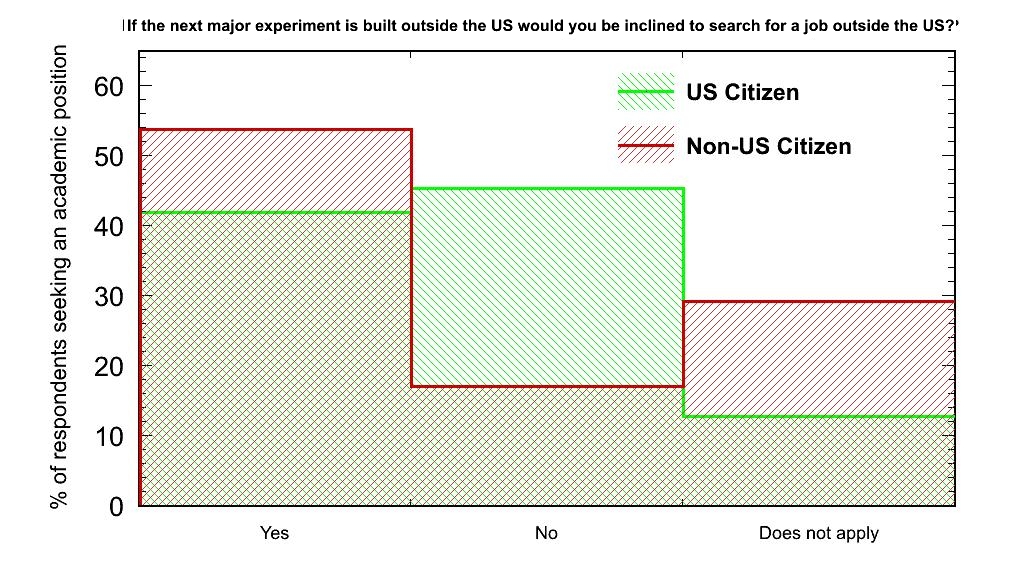}
\caption{Response from those who intend to seek an academic career to the question if they would be inclined to search for a job outside the U.S. if the next major experiment in their field was built outside the U.S. broken down by those who are U.S. citizens and those who are non-U.S. citizens.}
\label{fig:LeavingUSCitizen}
\end{center}
\end{figure}

\begin{figure}[htp]
\begin{center}
\includegraphics [scale=0.37]{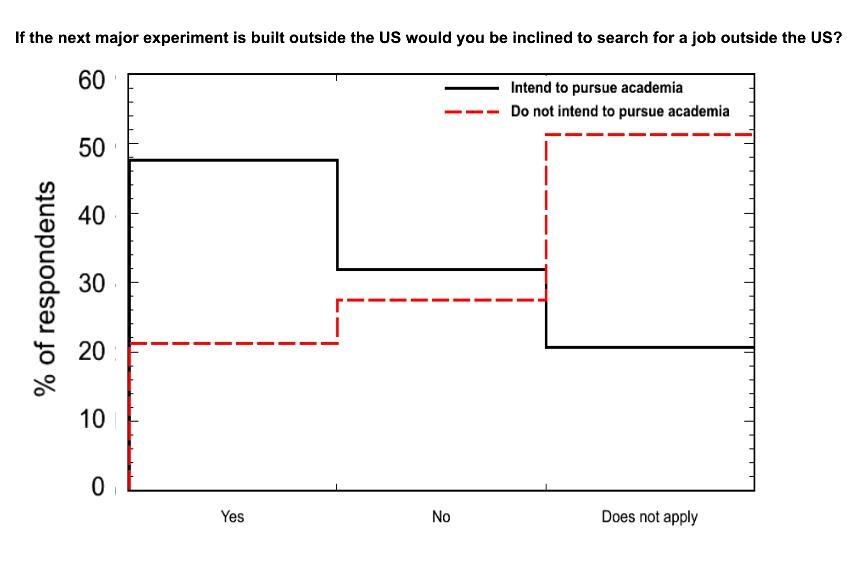}
\caption{Responses to the question if the survey participant would be inclined to search for a job outside the U.S. if the next major experiment in their field was built outside the U.S. broken down by those who intend to search for a job in academia.} \label{fig:LeavingUSAcademia}
\end{center}
\end{figure}

To follow-up on this observed trend we ask respondents where they will be more inclined to search for a job if the next major experiement in their frontier is built outside of the U.S.  Figure \ref{fig:LeavingUSAll} indicates that nearly 50$\%$ would be more likely to search for a job outside the U.S in this if the next major experiment is built outside of the U.S. This trend is greater amongst the non-U.S. members of HEP, as shown in Figure \ref{fig:LeavingUSCitizen}, and much more heavily present amongst those seeking a job in academia, as shown in Figure \ref{fig:LeavingUSAcademia}. 

\begin{figure}[htp]
\begin{center}
\includegraphics [scale=0.33]{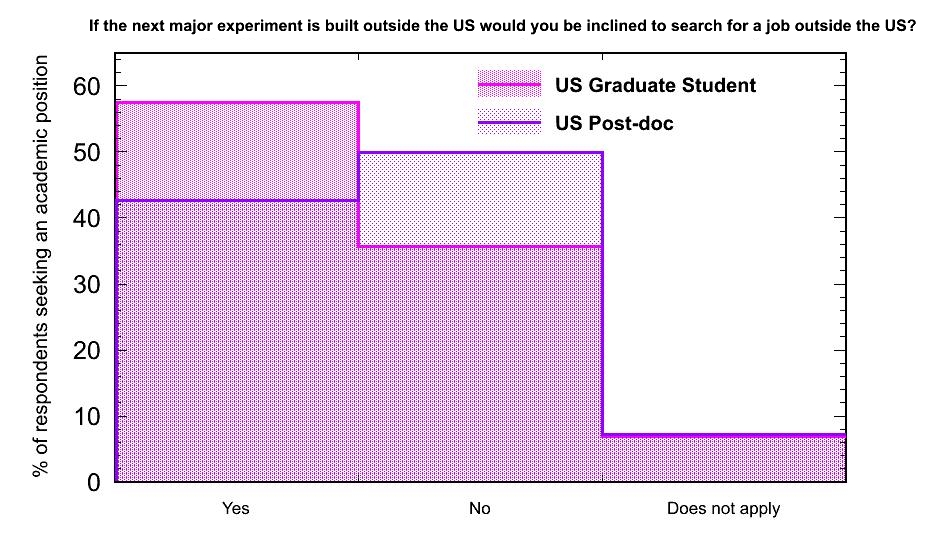}
\caption{Responses to the question if the survey participant would be inclined to search for a job outside the U.S. if the next major experiment in their field was built outside the U.S. broken down by those who are currently U.S. graduate students and U.S. Postdoc.}
\label{fig:LeavingGradPostdoc}
\end{center}
\end{figure}

\begin{figure}[htp]
\begin{center}
\includegraphics [scale=0.30]{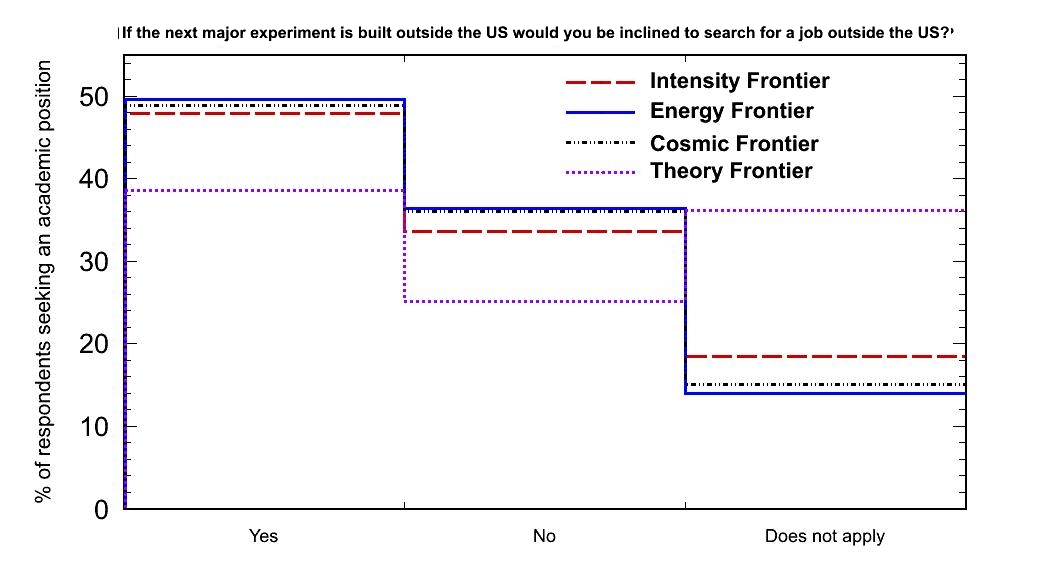}
\caption{Responses from those who intend to seek an academic career to the question if they would be inclined to search for a job outside the U.S. if the next major experiment in their field was built outside the U.S. broken down by various frontiers.}
\label{fig:LeavingUSFrontier}
\end{center}
\end{figure}

Finally, Figure \ref{fig:LeavingGradPostdoc} indicates that graduate students (the youngest in our field) are more inclined than post-docs to search elsewhere if the next major experiment is built outside the U.S. furthering the trends we observe above. Figure \ref{fig:LeavingUSFrontier} reinforces this and shows that those in the Intensity and Energy Frontier are more apt to search for a job outside the U.S. if the next major experiment from their frontier is built outside the U.S.

Taking these two questions together leads to the observation that while most of the respondents to the survey would prefer to seek a job in the U.S., there could be ``brain drain'' depending on whether or not major experiments take place in the U.S. This effect is particularly apparent among graduate students and the U.S.'s ability to attract non-U.S. scientists, but uniformly impacts all of the frontiers.

\subsection{Spires Data} \label{sec:spires}

To expand on these job related concerns further, we gathered some data from the INSPIRE High Energy Physics information system \cite{Spires:Data}. This data can be found at \url{http://hoc.web.cern.ch/hoc/jobs_stats2.txt}. 

\begin{figure}[htp]
\begin{center}
\includegraphics [scale=0.38]{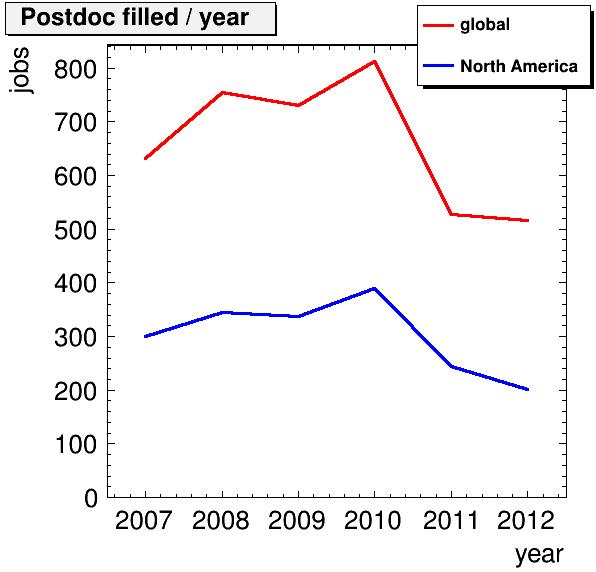}
\includegraphics [scale=0.38]{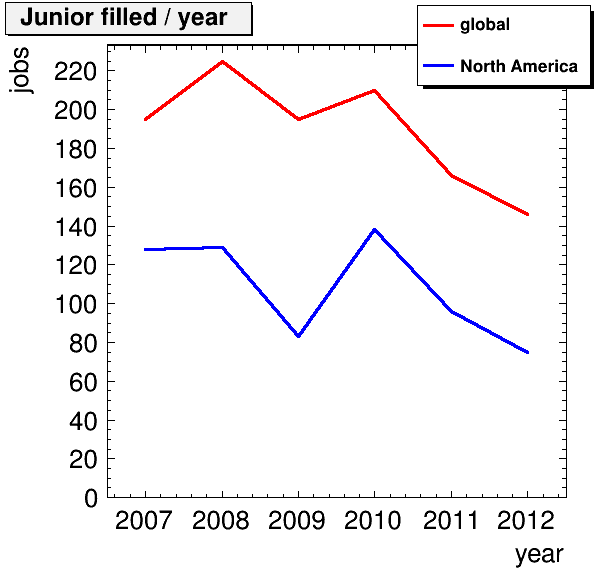}
\includegraphics [scale=0.38]{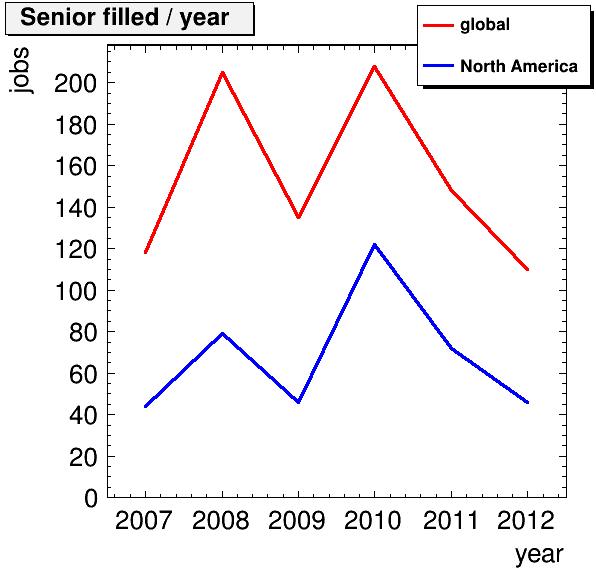}
\caption{Information gathered from the INSPIRE High Energy Physics information system about the available and filled jobs for postdocs, junior, and senior scientists from 2007 - 2012.}
\label{fig:Spires}
\end{center}
\end{figure}

Figure \ref{fig:Spires} shows the jobs listed on INSPIRE that were filled from 2007 - 2012. These jobs fall into the category of post-doc, junior, and senior positions. This job data is broken down by location, showing jobs listed in North America and jobs everywhere else (global). A clear downward trend in the last few years can be observed broadly in all markets, with no sustained growth whatsoever since 2007. 

A continuation of these trends will almost certainly lead to fewer young scientists remaining in the field of particle physics and exasperate the trends of failing to attract non-U.S. scientists, as indicated in the previous section.  

\section{Physics Outlook}\label{sec:physicsOutlook}

Questions in this section highlighted the science that is being planned during Snowmass 2013. We asked survey participants such questions as: which frontier will have the greatest impact in the next 10 years on HEP, to indicate which of the planned experiments (given from a non-exhaustive list) did the they have the highest priority for, and whether or not they would encourage other people to pursue the science within their frontier.

Some general trends were observed about the physics outlook of those in HEP responding to our survey:

\begin{enumerate}
\item \textbf{Would you encourage other young physicists to pursue a career in your frontier?}

More than 75$\%$ of the respondents would recommend other talented young physicists to pursue a career in their frontier. This is a trend that is shown to be true across all the frontiers and for both young and senior members of HEP. This particular fact is remarkable given the rather pessimistic outlook for funding and jobs and demonstrates that the science is found to be very compelling.

\begin{figure}[htp]
\begin{center}
\includegraphics [scale=0.31]{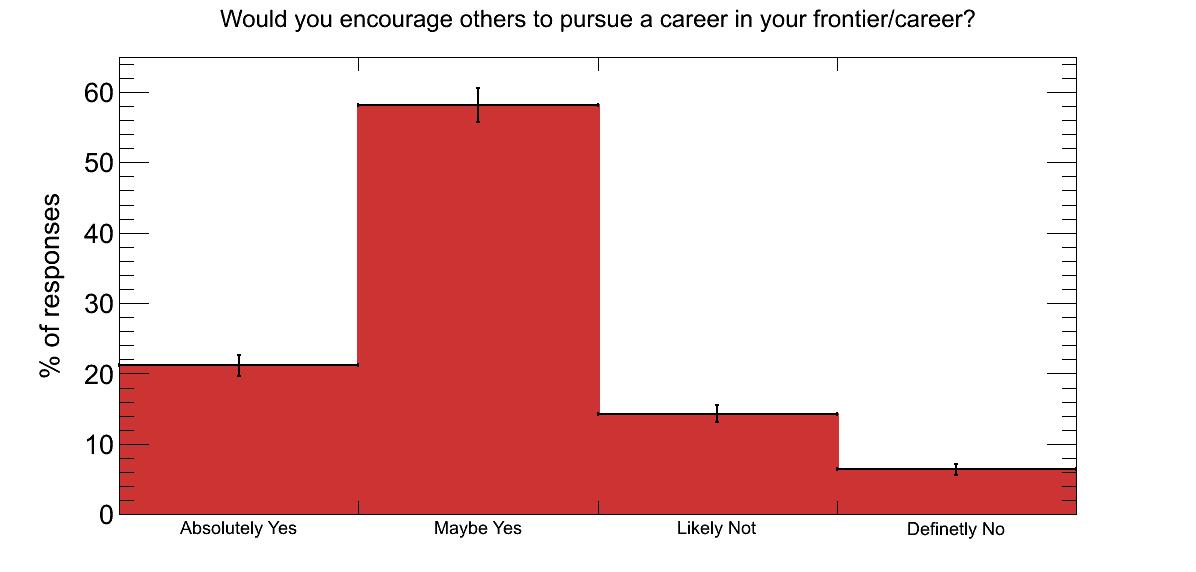}
\caption{Response to question about whether the participant would encourage someone to pursue a career in their frontier.} \label{fig:EncourageAll}
\end{center}
\end{figure}

\begin{figure}[htp]
\begin{center}
\includegraphics [scale=0.20]{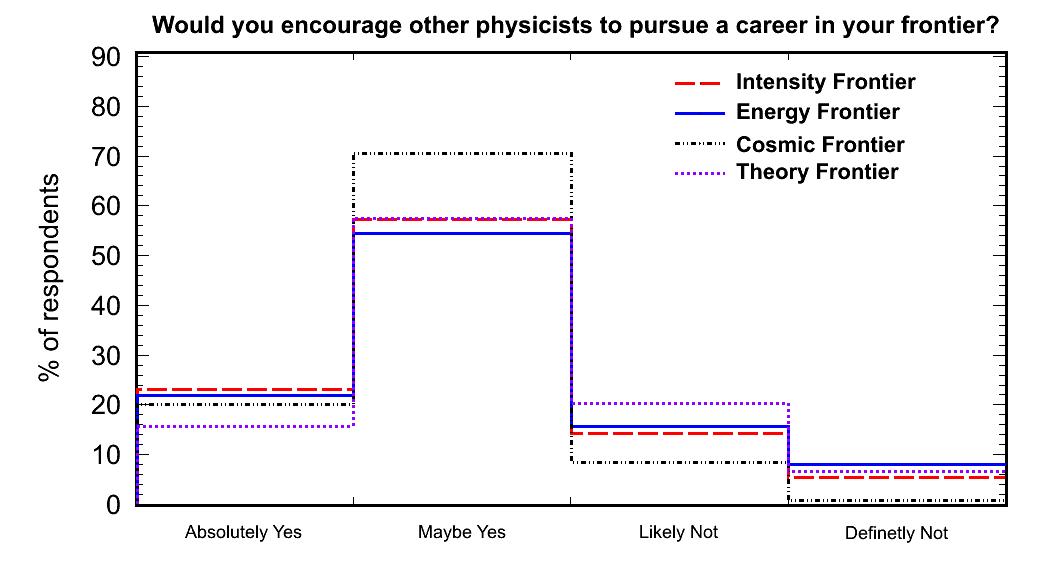}
\includegraphics [scale=0.20]{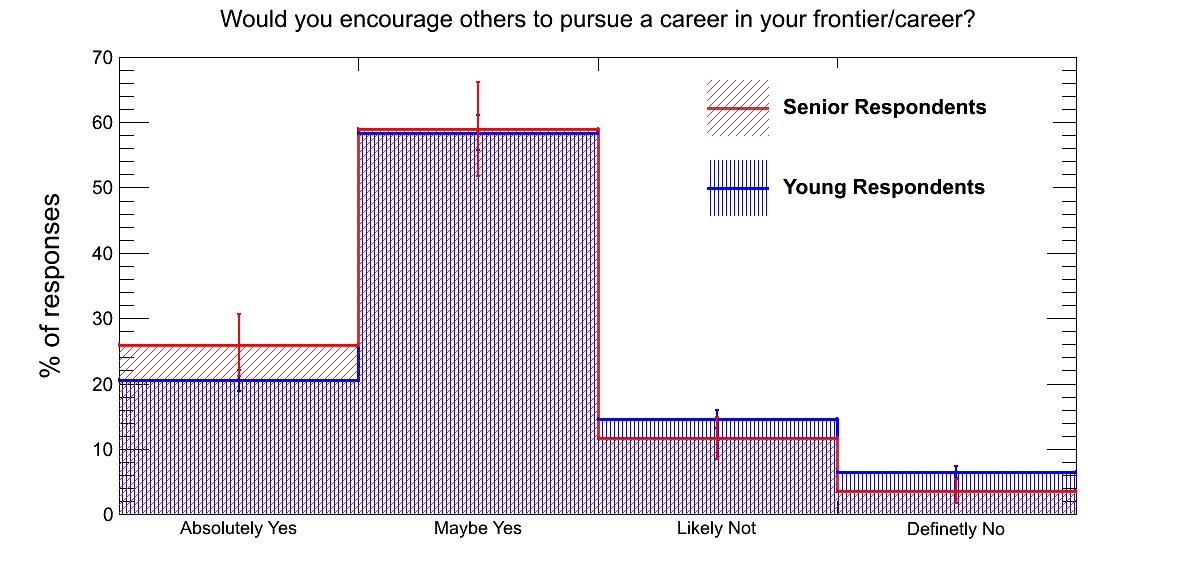}
\caption{Response to question about whether the participant would encourage someone to pursue a career in their frontier broken down by frontier and by whether the participant was young or senior.} \label{fig:EncourageBrokenDown}
\end{center}
\end{figure}

\item \textbf{Which of the following frontiers as defined by the Snowmass process will have the greatest impact on the landscape of High Energy Physics in the next 10 years?}

The results for this question are shown in Figures \ref{fig:ImpactonHEPAll} and \ref{fig:ImpactonHEPYounOld}.
The Energy frontier is seen as likely to have the greatest impact in the coming decade. However this fact seems less important than when broken down by frontier of the respondent. Our sample contained more Energy Frontier people, so we received more energy responses. Cosmic and Intensity frontiers appear in second and third position, respectively. Although the trend is the same for both groups, we notice that young scientists mentioned Energy frontier and Instrumentation less often than seniors, while preferring Cosmic, Computing and Theory slightly more often.

\begin{figure}[htp]
\begin{minipage}[c]{.475\linewidth}
\begin{center}
\includegraphics [scale = 0.21]{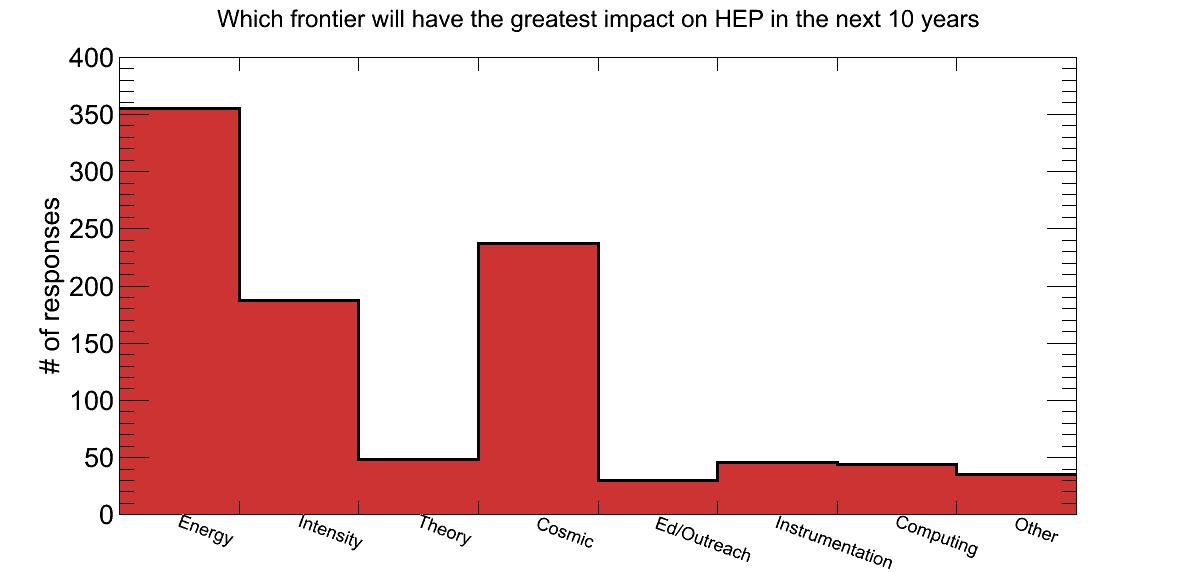}
\caption{Responses to which of the frontiers will have the greatest impact on HEP in the next 10 years.} \label{fig:ImpactonHEPAll}
\end{center}
\end{minipage}
\hspace{.05\linewidth}
\begin{minipage}[c]{.475\linewidth}
\begin{center}
\includegraphics [scale = 0.21]{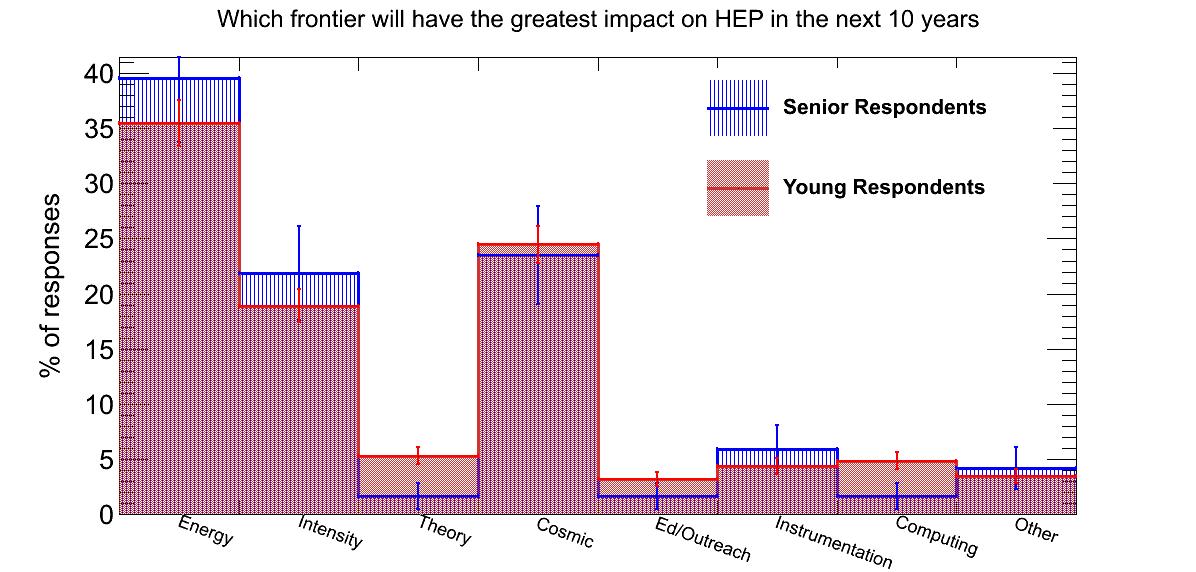}
\caption{Responses to which of the frontiers will have the greatest impact on HEP in the next 10 years broken down by young and senior definitions.}
\label{fig:ImpactonHEPYounOld}
\end{center}
\end{minipage}
\end{figure}

From these responses sub-samples were created based on which frontier people reported to be currently working on. Except for theorists, most respondents regarded their own current frontier as the one likely to have the greatest impact in the next 10 years (57\%,60\%,70\% for Intensity, Energy and Cosmic frontier, respectively). Theorists responded with the Energy Frontier (33\%) and Cosmic Frontier (30\%) likely to have the greatest impact.

\end{enumerate}

Another area we asked respondents was to indicate which of the planned experiments (given from a non-exhaustive list) from each of the three major frontiers they gave the highest priority to. Section \ref{sec:IntensitySub} shows the responses for Intensity Frontier experiments, Section \ref{sec:EnergySub} shows responses for the Energy Frontier, and Section \ref{sec:CosmicSub} shows the responses for the Cosmic Frontier. Each section is broken down by the current frontier the survey participant lists.

\subsection{Intensity Frontier Experiments}\label{sec:IntensitySub}

\begin{figure}[htp]
\begin{center}
\includegraphics [scale=0.40]{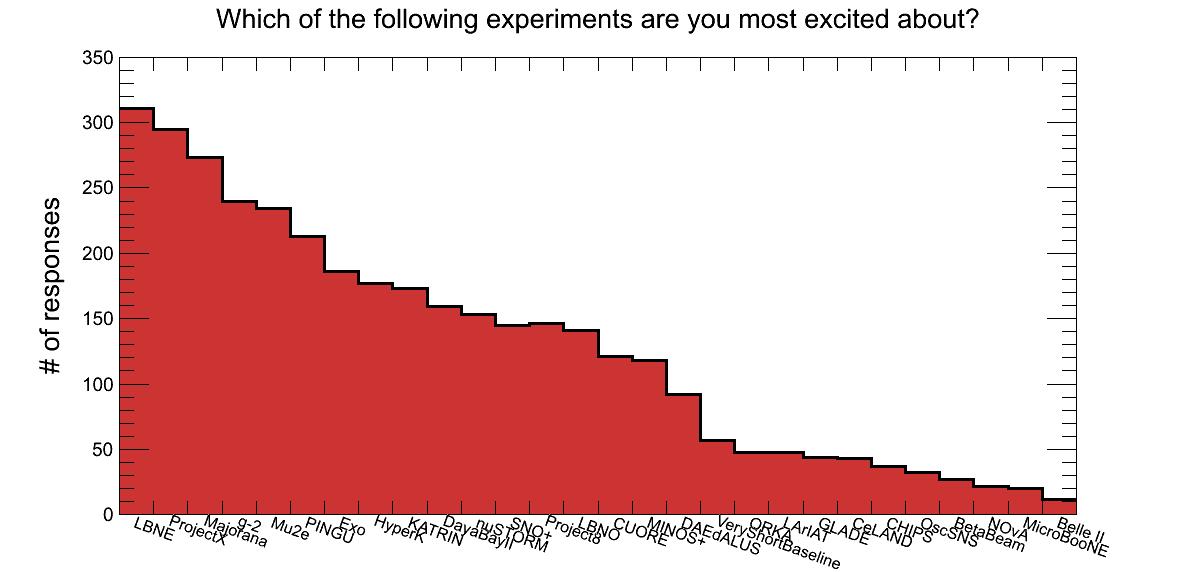}
\caption{The respondent was asked to select the most exciting experiments from the non-exhaustive list provided. The respondent could select more than one.} \label{fig:IntensityAll}
\end{center}
\end{figure}

Figure \ref{fig:IntensityAll} shows a non-exhaustive list of planned experiments from the intensity frontier which the survey respondent was asked to check which of the following experiments they are excited about. The respondent could select more than one experiment. The three intensity frontier experiments receiving the most overall votes are LBNE (312 votes), Project-X (295 votes), and Majorana (274 votes). 

We looked at which experiments that respondents from different frontiers were the most excited about. Figure \ref{fig:IntensityFrontierExp} shows the excitement for the various Intensity Frontier experiments broken down by the survey respondents current frontier. The top six Intensity Frontier experiments respondents were excited about, broken down by their current frontier, is shown in Table \ref{tab:IntensityExpFrontier}.

\begin{table}[H]
	\begin{center}
	\resizebox{\textwidth}{!}{%
	\begin{tabular}{|c|c|c|c|c|}
	\hline
	& \textbf{Cosmic Frontier} & \textbf{Theory Frontier} & \textbf{Energy Frontier} & \textbf{Intensity Frontier}\\
	\hline\hline
	1 & PINGU & Majorana & Project X & LBNE \\
	\hline
	2 & Majorana & g-2 & LBNE & Project X \\
	\hline
	3 & Exo & Mu2e & g-2 & nuStorm\\
	\hline
	4 & Sno+ & LBNE & Mu2e & PINGU\\
	\hline
	5 & Katrin & HyperK & Majorana & HyperK\\
	\hline
	6 & LBNE & Exo & Exo & Majorana\\
	\hline
	\end{tabular}}
	\caption{The top six Intensity Frontier experiments respondents were excited about, broken down by their current frontier.}\label{tab:IntensityExpFrontier}
	\end{center}
\end{table}

\begin{figure}[htp]
\begin{center}
\includegraphics [height=5.2cm,width=14cm]{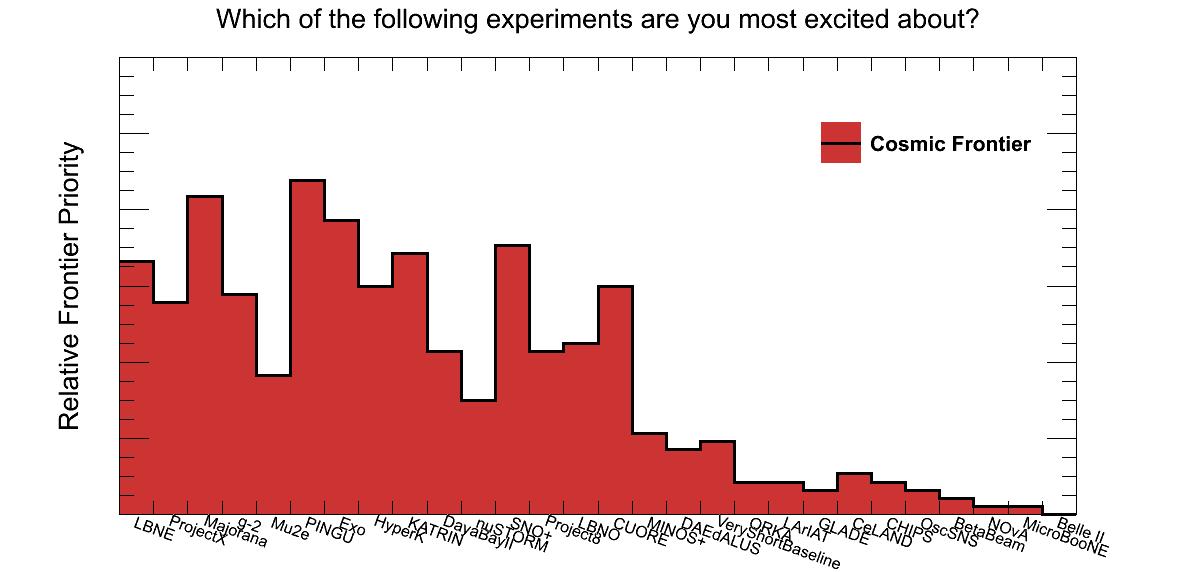}
\includegraphics [height=5.2cm,width=14cm]{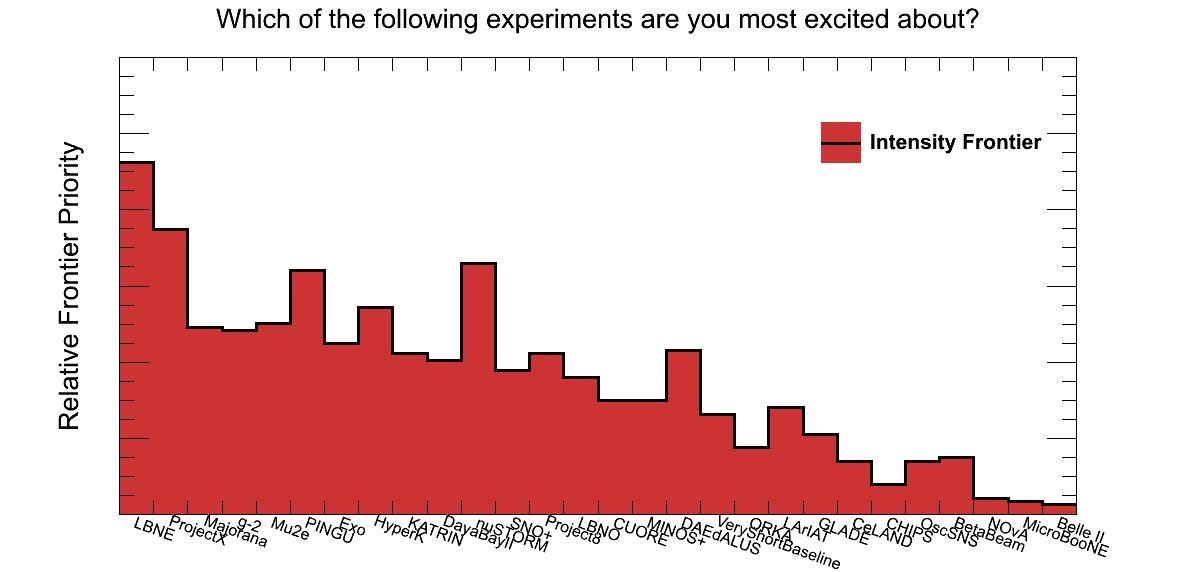}
\includegraphics [height=5.2cm,width=14cm]{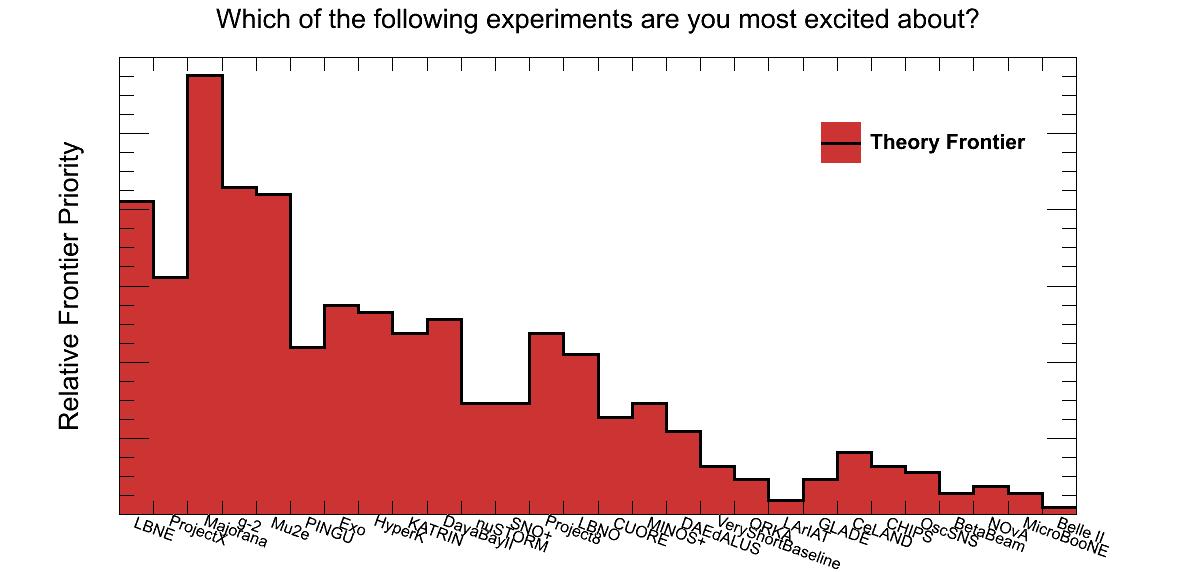}
\includegraphics [height=5.2cm,width=14cm]{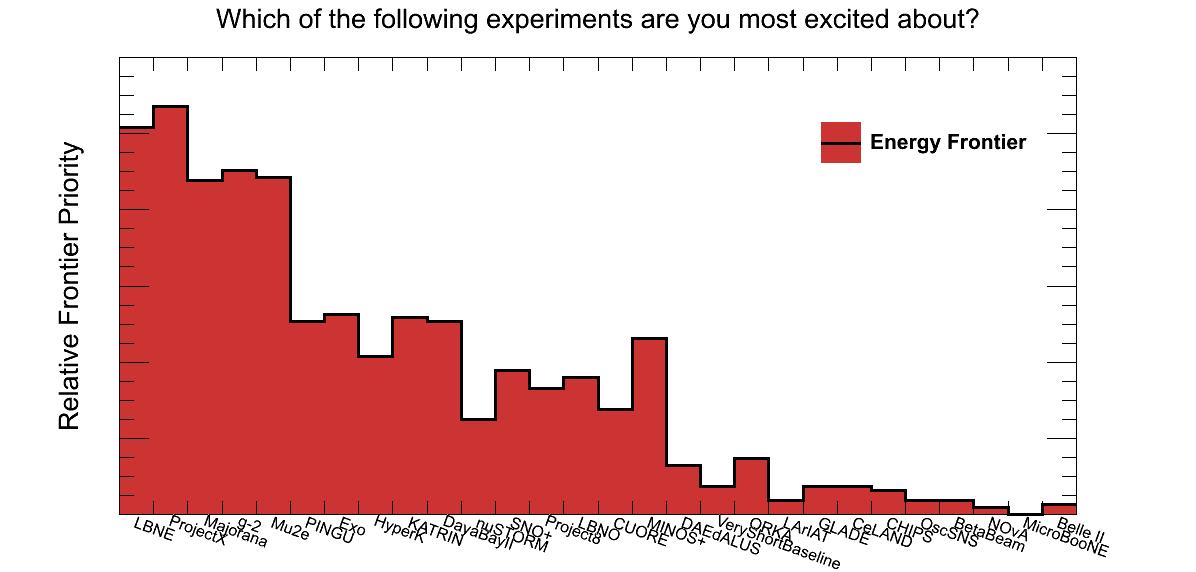}
\caption{The respondent was asked to select the most exciting experiments from the non-exhaustive list provided broken down frontier. The respondent could select more than one.} \label{fig:IntensityFrontierExp}
\end{center}
\end{figure}

\subsection{Energy Frontier Experiments}\label{sec:EnergySub}

\begin{figure}[htp]
\begin{center}
\includegraphics [scale=0.31]{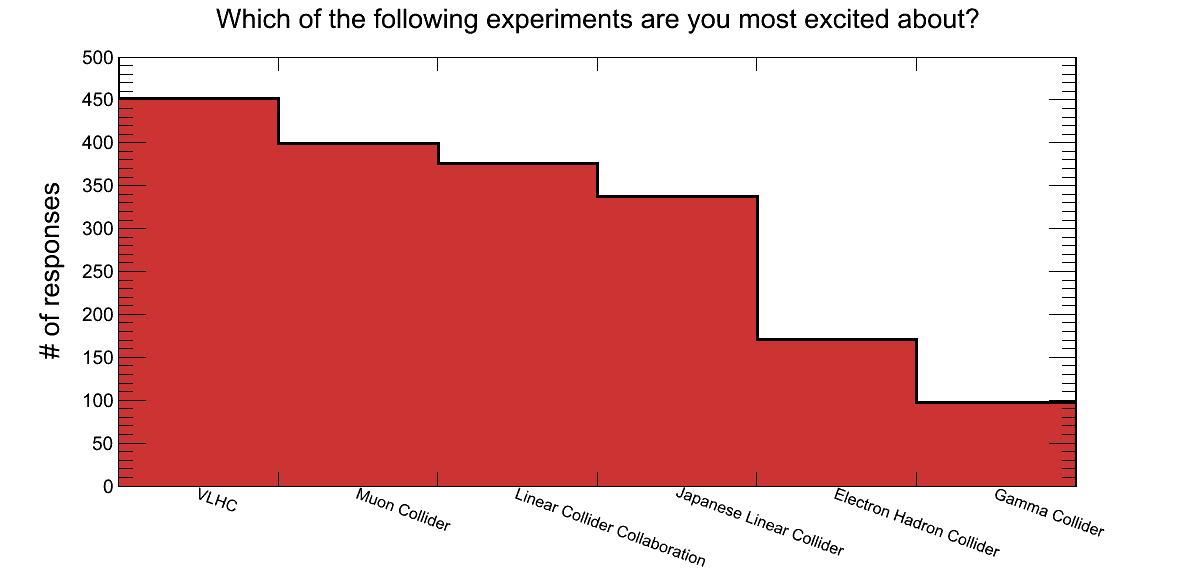}
\caption{The respondent was asked to select the most exciting experiments from the non-exhaustive list provided. The respondent could select more than one.} \label{fig:EnergyAll}
\end{center}
\end{figure}

Figure \ref{fig:EnergyAll} shows a list of planned experiments from the energy frontier which the survey respondent was asked to check which of the following experiments they are excited about. The respondent could select more than one experiment. The three energy frontier experiments receiving the most overall votes are VLHC (452 votes), Muon Collider (399 votes), and Linear Collider Collaboration (376 votes). 

Figure \ref{fig:IntensityFrontierExp} shows the excitement for the various Energy Frontier experiments broken down by the suvery respondents current frontier. The top six Energy Frontier experiments respondents were excited about, broken down by their current frontier, is shown in Table \ref{tab:EnergyExpFrontier}.

\begin{table}[H]
	\begin{center}
	\resizebox{\textwidth}{!}{%
	\begin{tabular}{|c|c|c|c|c|}
	\hline
	& \textbf{Cosmic Frontier} & \textbf{Theory Frontier} & \textbf{Energy Frontier} & \textbf{Intensity Frontier}\\
	\hline\hline
	1 & VLHC & VLHC & VLHC & Muon Collider \\
	\hline
	2 & Linear Collider Collaboration & Japanese Linear Collider & Linear Collider Collaboration  & Japanese Linear Collider \\
	\hline
	3 & Muon Collider & Linear Collider Collaboration & Muon Collider &  Linear Collider Collaboration\\
	\hline
	4 & Japanese Linear Collider & Muon Collider & Japanese Linear Collider & VLHC\\
	\hline
	5 & Electron Hadron Collider & Electron Hadron Collider & Electron Hadron Collider & Electron Hadron Collider\\
	\hline
	6 & Gamma Collider & Gamma Collider & Gamma Collider & Gamma Collider\\
	\hline
	\end{tabular}}
	\caption{The top six Energy Frontier experiments respondents were excited about, broken down by their current frontier.}\label{tab:EnergyExpFrontier}
	\end{center}
\end{table}

\begin{figure}[htp]
\begin{center}
\includegraphics [height=5.2cm,width=14cm]{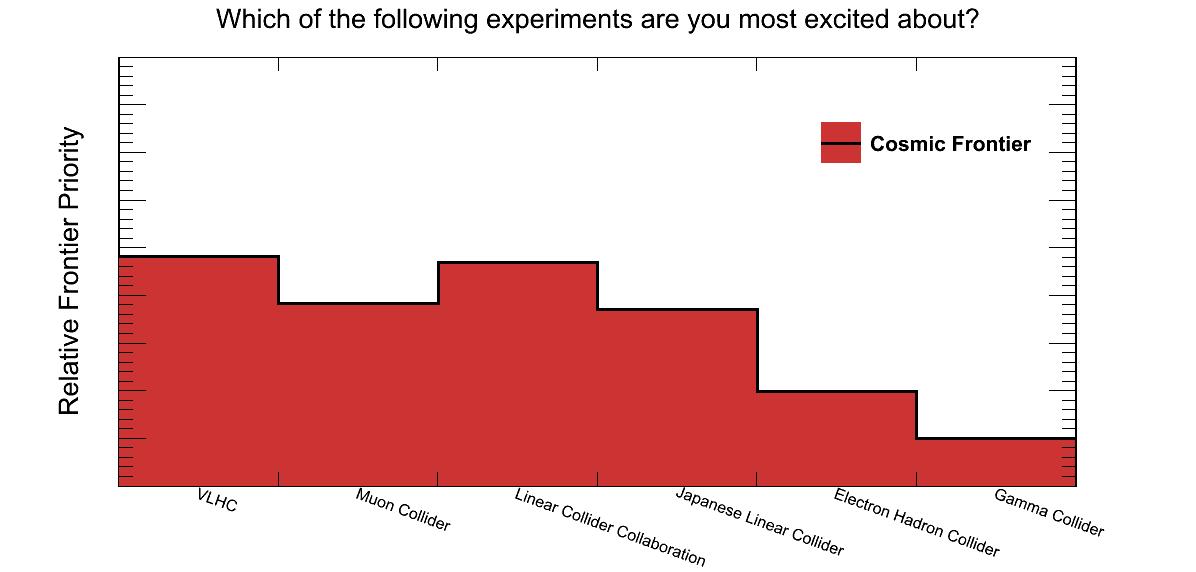}
\includegraphics [height=5.2cm,width=14cm]{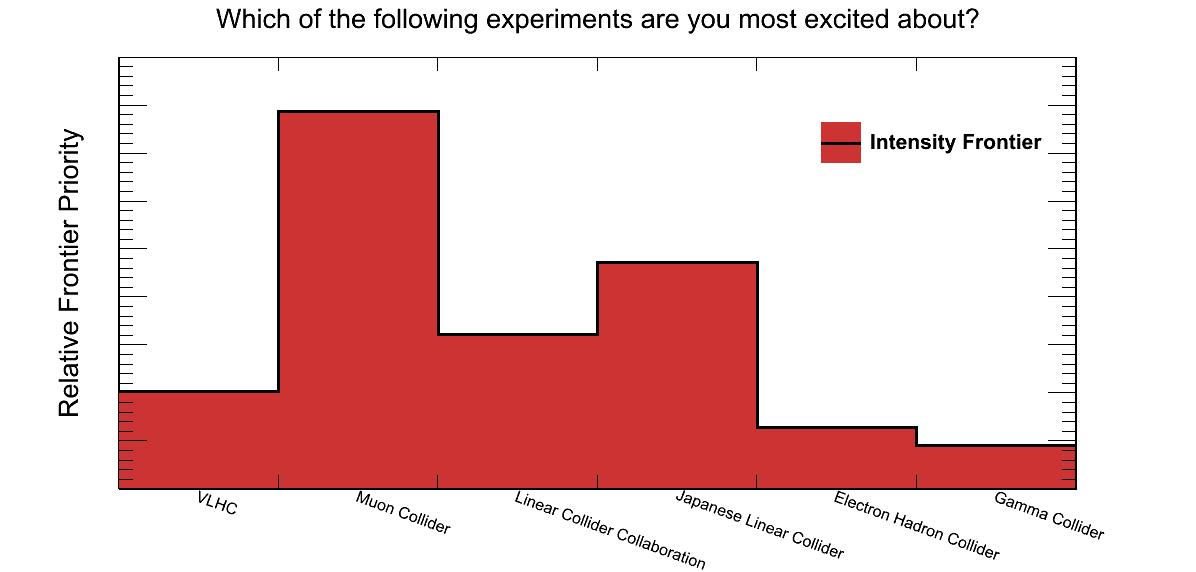}
\includegraphics [height=5.2cm,width=14cm]{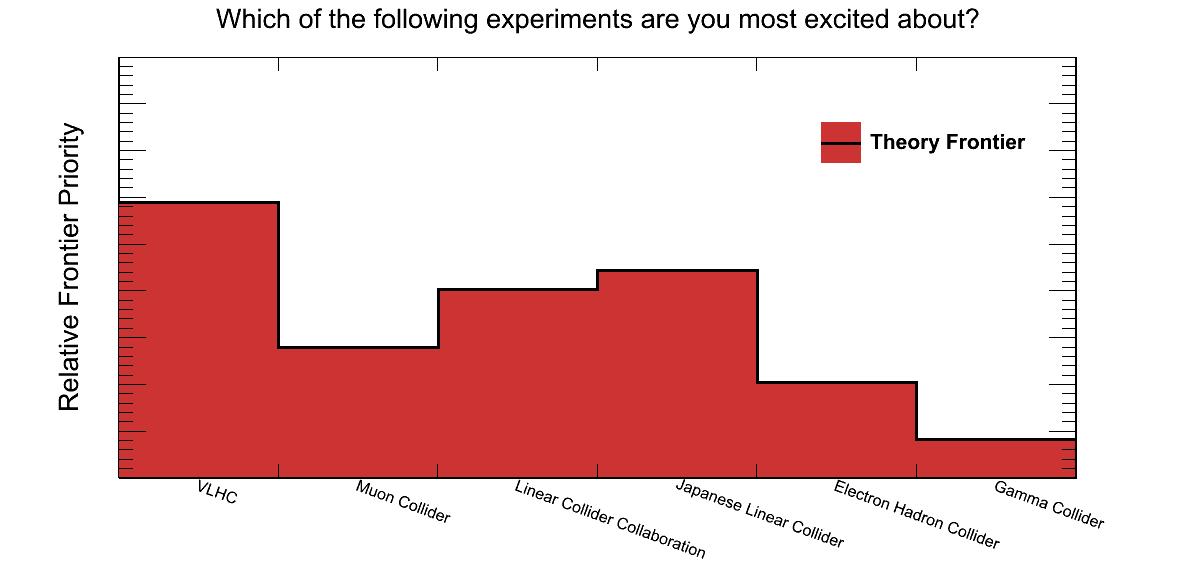}
\includegraphics [height=5.2cm,width=14cm]{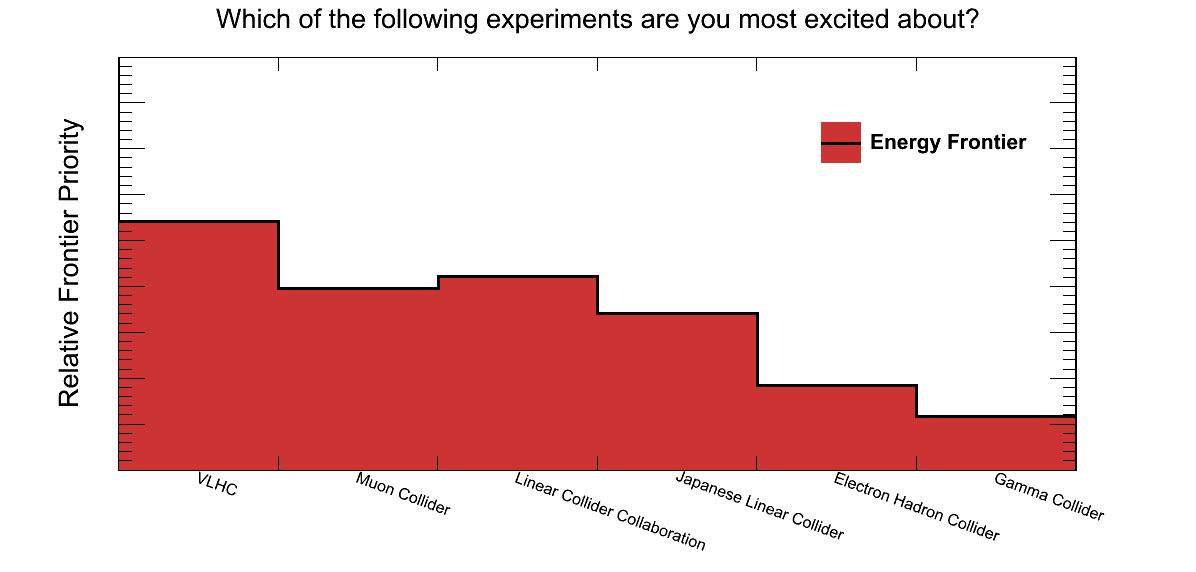}
\caption{The respondent was asked to select the most exciting experiments from the non-exhaustive list provided broken down by frontier. The respondent could select more than one.} \label{fig:IntensityFrontierExp}
\end{center}
\end{figure}

\subsection{Cosmic Frontier Experiments}\label{sec:CosmicSub}

\begin{figure}[htp]
\begin{center}
\includegraphics [scale=0.31]{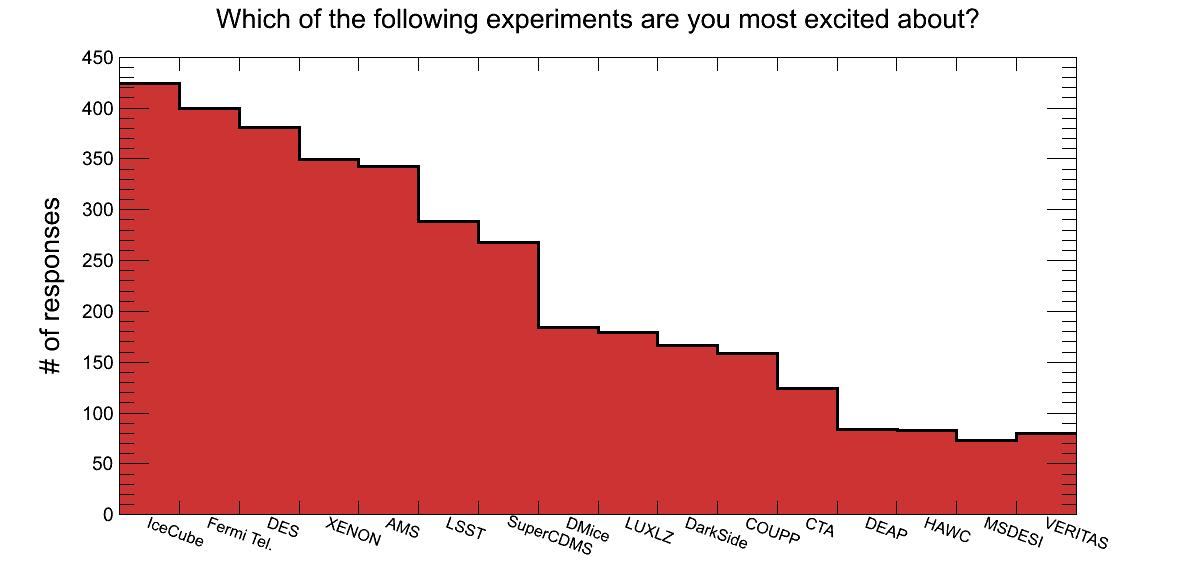}
\caption{The respondent was asked to select the most exciting experiments from the non-exhaustive list provided. The respondent could select more than one.} \label{fig:CosmicAll}
\end{center}
\end{figure}

Figure \ref{fig:CosmicAll} lists the planned experiments from the Cosmic Frontier which the survey respondent was asked to check which of the following experiments they are excited about. The respondent could select more than one experiment. The three cosmic frontier experiments receiving the most overall votes are IceCube (454 votes), Fermi Telescope (400 votes), and Dark Energy Survey (381 votes). 

\begin{figure}[htp]
\begin{center}
\includegraphics [height=5.2cm,width=14cm]{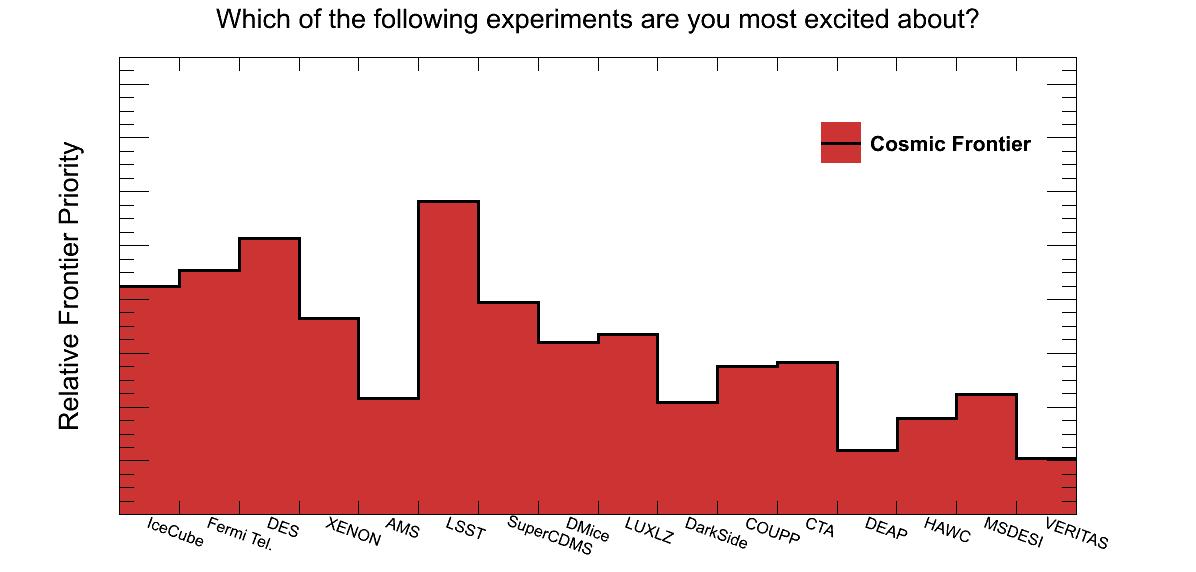}
\includegraphics [height=5.2cm,width=14cm]{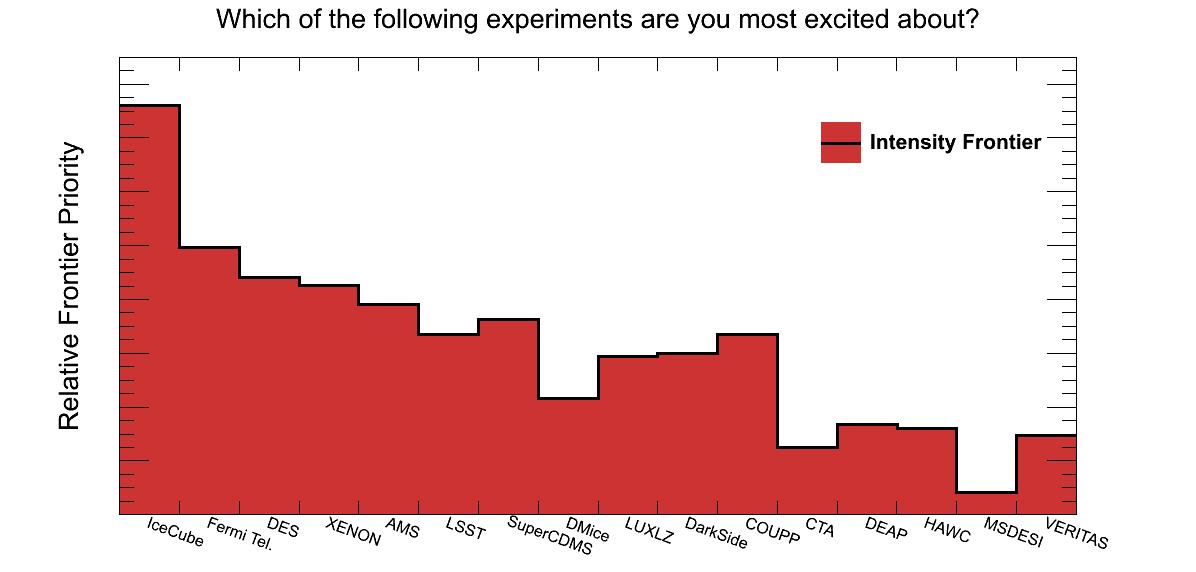}
\includegraphics [height=5.2cm,width=14cm]{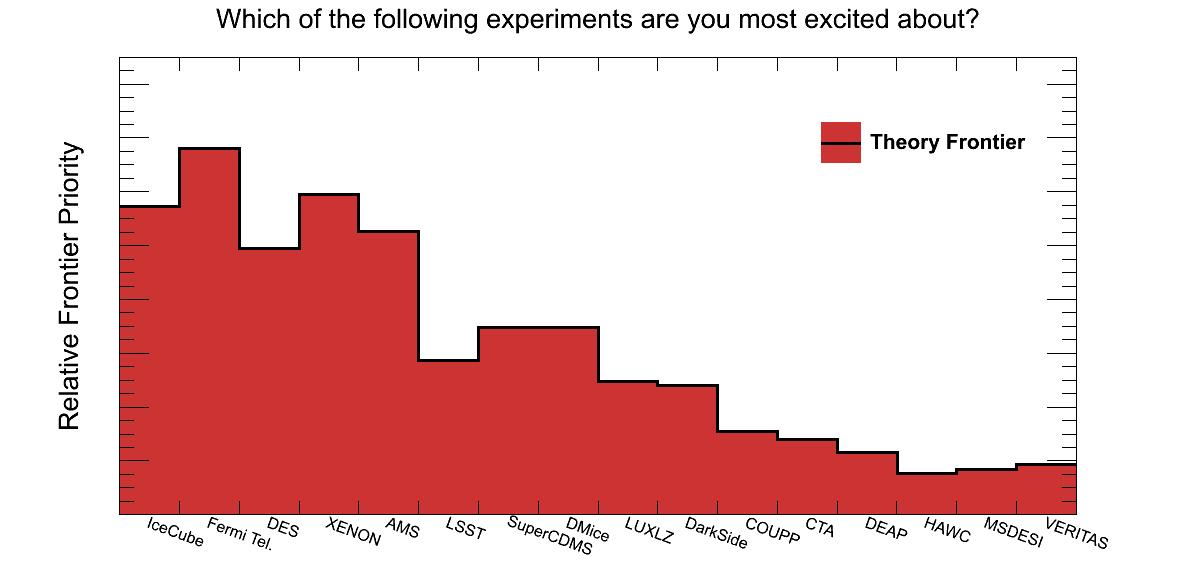}
\includegraphics [height=5.2cm,width=14cm]{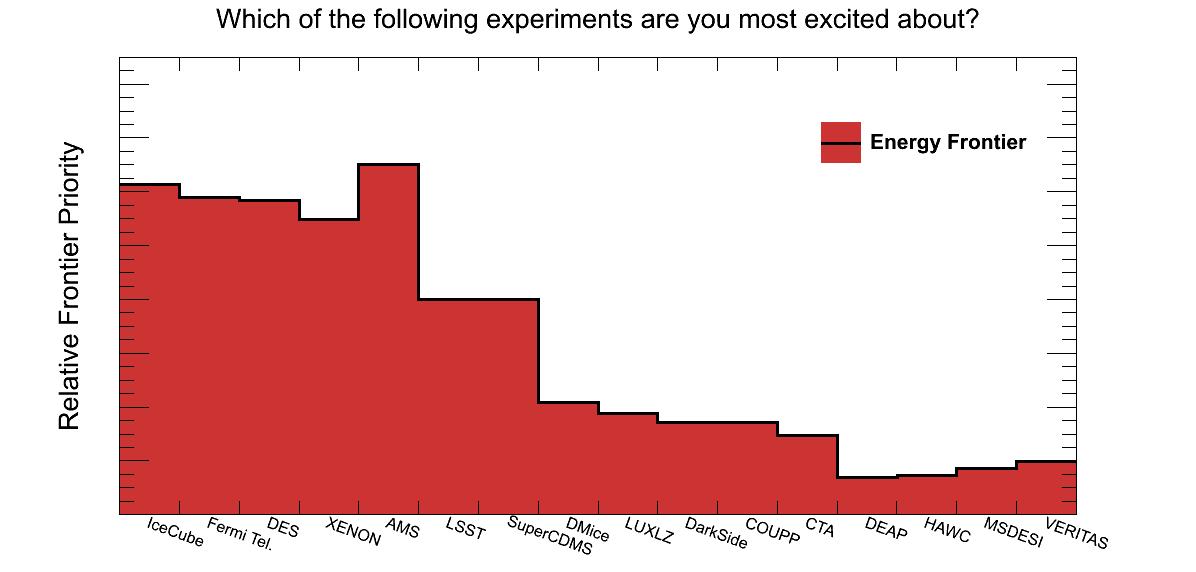}
\caption{The respondent was asked to select the most exciting experiments from the non-exhaustive list provided broken down by frontier. The respondent could select more than one.} \label{fig:CosmicFrontierExp}
\end{center}
\end{figure}

Figure \ref{fig:CosmicFrontierExp} shows the excitement for the various Cosmic Frontier experiments broken down by the suvery respondents current frontier. The top six Cosmic Frontier experiments respondents were excited about, broken down by their current frontier, is shown in Table \ref{tab:CosmicExpFrontier}.

\begin{table}[H]
	\begin{center}
	\resizebox{\textwidth}{!}{%
	\begin{tabular}{|c|c|c|c|c|}
	\hline
	& \textbf{Cosmic Frontier} & \textbf{Theory Frontier} & \textbf{Energy Frontier} & \textbf{Intensity Frontier}\\
	\hline\hline
	1 & LSST & Fermi Telescope & AMS & IceCube \\
	\hline
	2 & Dark Energy Survey & XENON & IceCube  & Fermi Telescope \\
	\hline
	3 & Fermi Telescope & IceCube & Fermi Telescope &  Dark Energy Survey\\
	\hline
	4 & IceCube & AMS & Dark Energy Survey & XENON\\
	\hline
	5 & SuperCDMS & Dark Energy Survey & XENON & AMS\\
	\hline
	6 & XENON & SuperCDMS & LSST & SuperCDMS\\
	\hline
	\end{tabular}}
	\caption{The top six Cosmic Frontier experiments respondents were excited about, broken down by their current frontier.}\label{tab:CosmicExpFrontier}
	\end{center}
\end{table}

Clearly there is excitement about many of the upcoming experiments proposed through Snowmass. It is worthwhile remarking that across the frontiers, the experiment seen from within the frontier as having the highest priority never has the highest priority across the other three frontiers. Taking the example of the Energy frontier, the VLHC is clearly seen as exciting and a high priority within the Energy frontier as well as with the Cosmic and Theory frontier. However, this experiment receives only a third rank within the Intensity frontier. Similar trends can be said of the Intensity frontier and the long baseline experiments as well as the Cosmic frontier and the LSST. Seeing the priorities vary so much across frontiers indicates that much more work is required to build consensus in the coming months.

\section{Non-academic career paths} \label{sec:NonAcademic}

In addition to our survey reaching those currently working in HEP, we endeavoured to reach people who had received their training within HEP but have gone on to pursue careers outside of academia. In particular we asked how people on non-academic career paths work lives compare with people pursuing an academic career. In total we received 74 responses from people now working outside of HEP.

In an effort to expand this sample, a handout was provided at Snowmass collecting information from senior scientists about former colleagues who have since left the field. This form can be found at \url{http://snowmassyoung.hep.net/OA_contacts.pdf}. We urge any reader who may have contact information to please visit this form and email \url{snowmass2013young@gmail.com} with the information.

 Some general trends we observed from this group responding to our survey:

\begin{enumerate}
\item \textbf{Before going into your current field of work, did you attempt to find a job in academics?}

57$\%$ of the survey respondents did not attempt to find a job inside academia before pursuing their career.

\begin{figure}[htp]
\begin{center}
\includegraphics [scale=0.30]{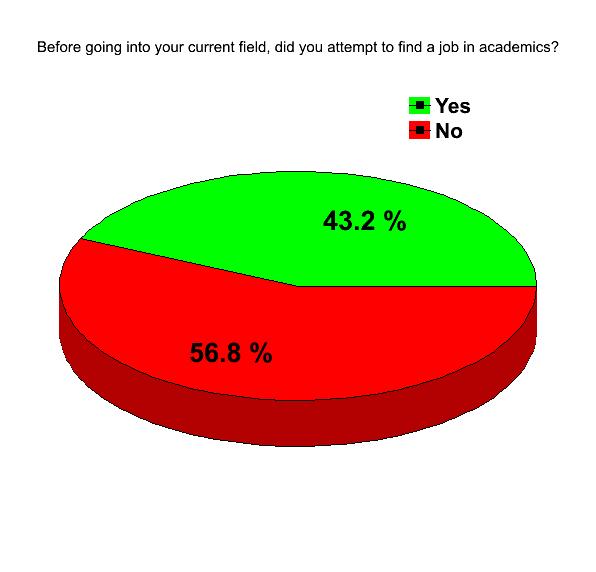}
\caption{Response from those on a non-academic career path as to whether or not they attempted to pursue a career in academia.}\label{fig:OAJobAttempt}
\end{center}
\end{figure}

\item \textbf{How many hours per week (weekend) do you work on average?}

Respondents from non-academic career paths report spending, on average, 50 hours per week at work, of which 7 hours are spent on the weekend working. This trend mirrors the respondents from academic career paths who report, on average, 49 hours per week at work, of which 7 hours are spent on the weekend working as well. This result ran counter to many of our intuitions about who works more both inside and outside of an academic career path.

\begin{figure}[htp]
\begin{center}
\includegraphics [scale=0.35]{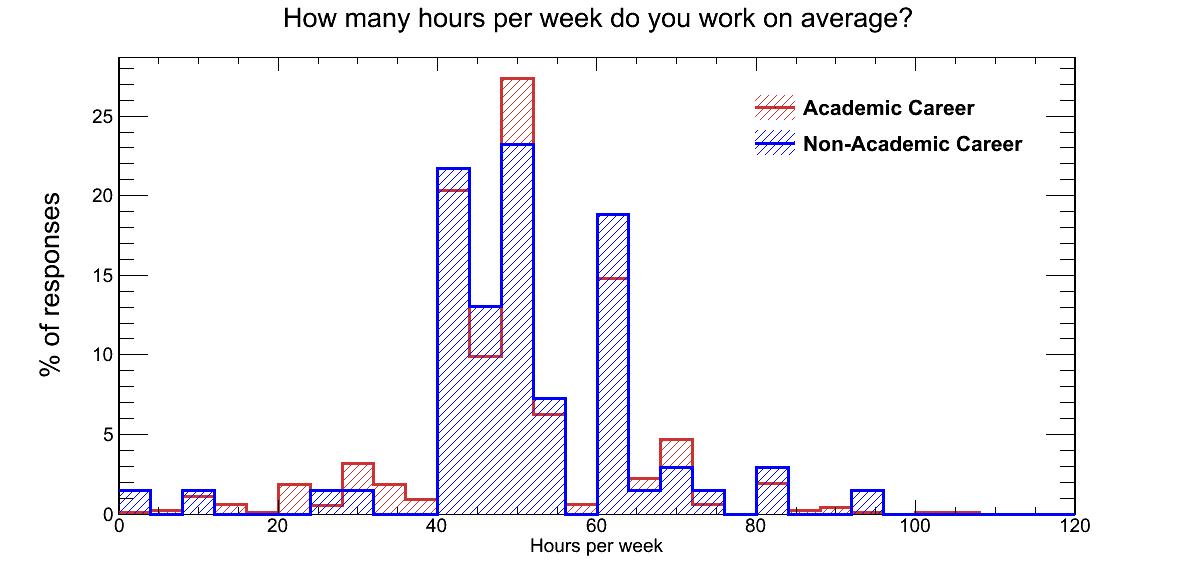}
\caption{Response from those on a non-academic career path as well as those on an academic career path to the average number of hours worked per week.}\label{fig:hoursworked}
\end{center}
\end{figure}

\item \textbf{In the last year, how many weeks have you had to travel for work related to your current position?}
The overwhelming majority of respondents from non-academic career paths report typically less than 4 weeks for work related travel in the last year. While the majority of respondents from academic career paths report spending greater than 4 weeks of work related travel. 

\begin{figure}[htp]
\begin{center}
\includegraphics [scale=0.35]{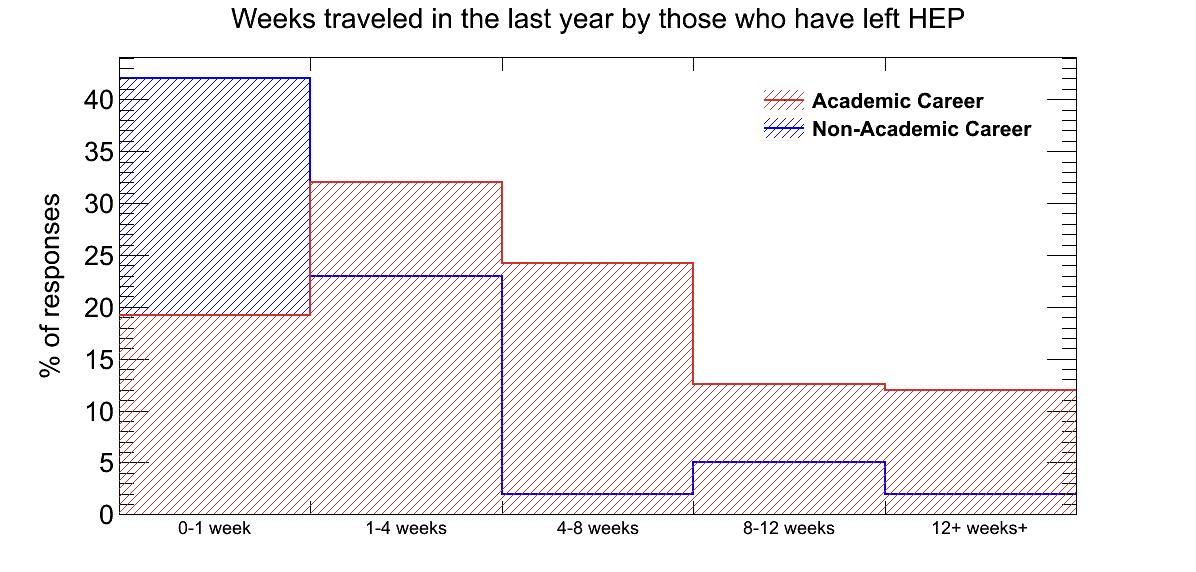}
\caption{Response from those on a non-academic career path as well as those on an academic career path to the number of weeks spent traveling for work in the last year.}\label{fig:hourstraveled}
\end{center}
\end{figure}

\end{enumerate}

\begin{figure}[htp]
\begin{minipage}[c]{\linewidth}
\begin{center}
\includegraphics [scale = 0.35]{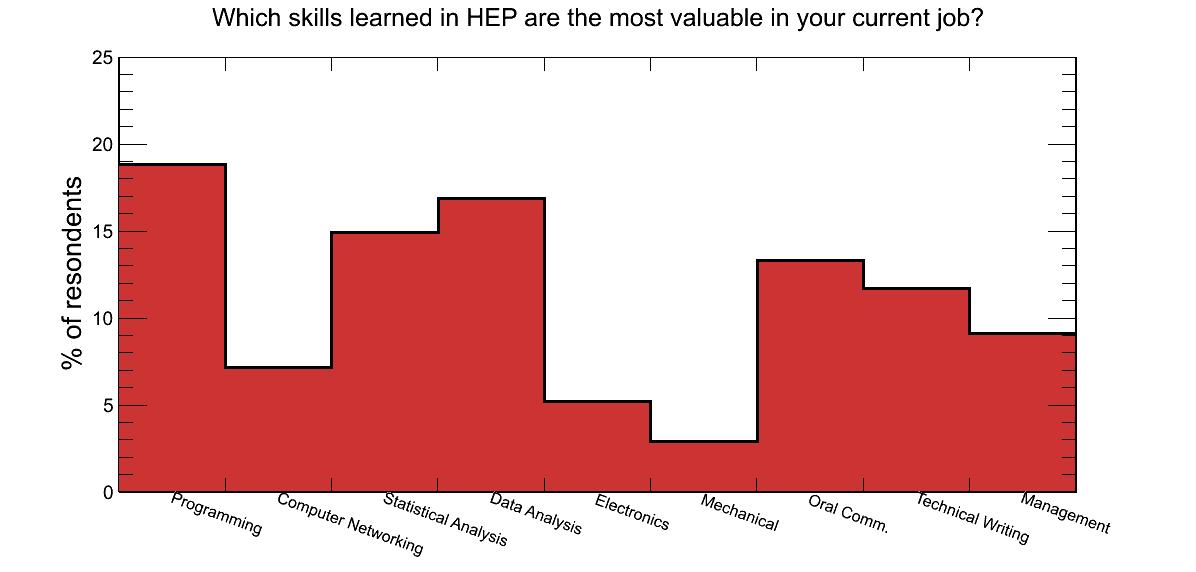}
\caption{List of skills learned during the respondents experience in HEP that they found valuable in their current position.}\label{fig:OASkills}
\end{center}
\end{minipage}
\end{figure}

Figure \ref{fig:OASkills} shows the main skills that people found most valuable to their current career include programming, data analysis, and statistical analysis. It is also worth noting that skills such as oral communication and technical writing are seen as almost as important in their current job.

\begin{figure}[htp]
\begin{minipage}[c]{\linewidth}
\begin{center}
\includegraphics [scale = 0.35]{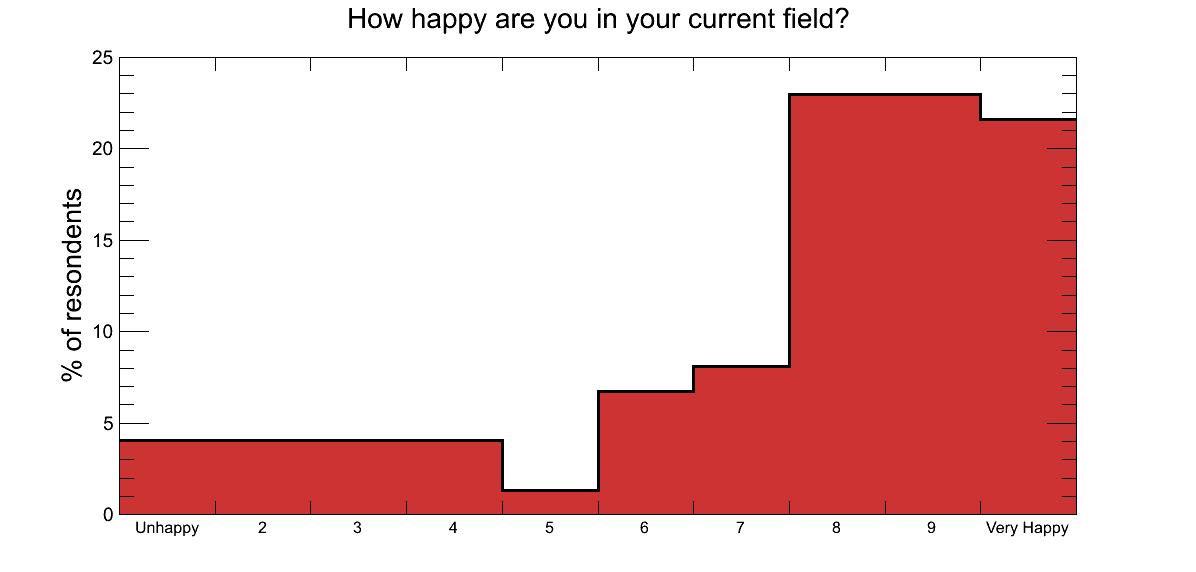}
\caption{On a scale of 1-10, how happy the non-academic career path respondent is in their current field.}\label{fig:OAHappy}
\end{center}
\end{minipage}
\end{figure}

Finally, Figure \ref{fig:OAHappy} shows that the overwhelming majority of the non-academic career path respondents are very happy with their current field of employment. 

\section{Conclusion}
The survey administered by the SYP was intended to collect a range of opinions that reflect the physics interests, the career outlook, and the general mood of the field leading up to Snowmass 2013, we collected 1112 total responses, 74 of which came from people who had received their training from HEP and have since chosen to pursue a non-academic career path.

Some broad conclusions we would like to draw from a look at the survey data:

\begin{enumerate}
\item[I] \textbf{Demographic Conclusion:}

The survey contained a large fraction of young people in HEP, many of whom had not been participating in Snowmass prior to taking this survey and many of which did not plan on attending the Snowmass meeting. This makes the viewpoints and opinions expressed even more important as it is likely their voices would go unheard otherwise.

\item[II] \textbf{Career Outlook Conclusion:}
Despite the wide spread perspective that the funding situation in the next ten years will likely be bleak, most young people are excited about the prospect of pursuing a job within HEP. Moreover, many of the respondents, both U.S. citizen and non-U.S. citizen intend to pursue a job within the U.S. . However, all this can shift if the U.S. misses an opportunity to build the next major physics experiment most relevant to the various frontiers in HEP, helping support the idea that the most compelling science will attract the best and brightest minds.

\item[III] \textbf{Physics Outlook Conclusion:}
There is a general sense of excitement about the science and the upcoming experiments being proposed at Snowmass. This excitement is seen across the frontiers and is shared by both the more senior members as well as the young scientists. However, the breakdown of which experiments that people from other frontiers find exciting shows there is still a lot of work left to do to build consensus.

\item[IV] \textbf{Non-academic Career Path Conclusion:}
The self reported work habits seem the mostly the same for both academic and non-academic careers. Furthermore, many of the skills learned in HEP are seen as valuable skills for those on a non-academic career path. Finally, those who have received their training in HEP and are now pursuing careers outside of academia are generally very happy with their current careers.

\end{enumerate}

\newpage


\newpage

\section{Survey Questions}\label{sec:surveyquestions}

\subsection*{\textbf{Demographic}}
\begin{itemize}
\item{What is your gender?}
\item{What is your current marital status?}
\item{Are you a US citizen?}
\item{What is the most advanced degree you currently hold?}
\item{How many children do you have?}
\item{What is your household salary (USD/year)?}
\item{In what country did you earn your most recent degree?}
\item{At what type of institution are you currently employed?}

\end{itemize}
\subsection*{Non-academic Career Path}

\begin{itemize}

\item{What is your current field of employment?}
\item{How long have you been in your current position?}
\item{In the last year, how many weeks have you had to travel for work related to your current position}
\item{Before going into your current field of work, did you attempt to find a job in academics?}
\item{What resources did you find useful in finding your first non-academic job?}
\item{Have you ever not pursued a job offer because it would require relocating?}
\item{Do you maintain a residence in a different location than your spouse/partner because of you work?}
\item{Have you found your career options limited due to personal relationships?}
\item{On a scale of 1 to 10, how well did your HEP physics experience prepare you for your current job?}
\item{What skills learned during your HEP physics experience were the most valuable to you in your current job?}
\item{Would you encourage other young physicists to pursue a career in your field?}
\item{On a scale of 1 to 10 how happy are you with your decision to work in your current field?}
\item{Which of the following would you rather be doing now in your career?}

\end{itemize}

\subsection*{Academic Career Outlook}

\begin{itemize}
\item{Please choose which category best describes your current position}
\item{How long have you been in your current category?}
\item{According to the Snowmass definitions, which froniter best describes the research you primarily work on?}
\item{According to the Snowmass definitions, which frontier do you see yourself working on 5 years from now?}
\item{Which of the following institutions do you do the majority of your research at?}
\item{Have you been contributing to the pre-Snowmass meetings and/or the Snowmass process prior to taking this survey?}
\item{Do you plan on attending the Snowmass on the Mississippi in Minneapolis, MN in July 2013?}
\item{Do you maintain a residence in a different location than your spouse/partner or children in order to work or study?}
\item{During your current position, which of the following best describes your primary responsibility?}
\item{In the last year, how many weeks have you had to travel for work related to your current position?}
\item{Which of the following best describes your primary reason for your work related travel?}
\item{In the last year, how many conference talks did you give?}
\item{Select two institutions which you most desire to be employed with after completing your training in High Energy Physics}
\item{Do you intend to pursue a permanent career in academia?}
\item{What do you think are the odds that you will obtain a permanent academic position?}
\item{What percentage of postdocs in HEP do you think go on to permanent positions in HEP?}
\item{How much consideration have you given to pursuing a job outside of academia?}
\item{In the future, will you be searching for a permanent position in the US or abroad?}
\item{If the next major expreiment in your frontier was built outside the US, would you be more inclined to search for a permanent position outside the US?}
\item{Would you encourage other young physicist to pursue a career in your frontier?}
\item{Have you found your career options limited due to personal relationships?}
\item[]{\textbf{Please grade each of the following career related concerns in terms of which you find the most important to you and your future in high energy physics.}}
\item{Availability of university based jobs}
\item{Availability of laboratory based jobs}
\item{Funding for large scale long lead time experiments in the future}
\item{Funding for small scale short lead time experiments in the future}
\item{Proximity to your home institution of future planned experiments}
\item{Bureaucracy and administrative difficulties to conduction research}

\end{itemize}

\subsection*{Physics Outlook}
\begin{itemize}
\item{On a scale of 1 to 10 (1 = Funding will stop, 10 = Funding will thrive) how do you feel about the funding within your frontier within the next decade}
\item[]{\textbf{Which of the following statements do you most agree with}}
\item{I believe we should invest the majority of our resources in one big project next (e.g. Linear Collider, LSST, Long Baseline Neutrino Experiment, Muon Collider, VLHC, etc...)}
\item{I believe we should invest the majority of our resources in a larger variety of smaller experiments (e.g. Mu2e, Project 8, LAr1, MS-DESI etc...)}
\item[]{}
\item{Which of the following frontiers as defined by the Snowmass process will have the greatest impact on the landscape of High Energy Physics in the next 10 years?}
\item[]{\textbf{Pick from a list of future planned experiments which you are most excited about:}}
\item{Intensity Frontier Experiments}
\item{Energy Frontier Experiments}
\item{Cosmic Frontier Experiments}

\end{itemize}
\subsection*{Concluding Questions}
\begin{itemize}
\item{How many hours per week do you work (on average)?}
\item{How many hours per weekend do you work (on average)?}
\item{How many hours per week do you spend in meetings (on average)?}
\item{Do you think your training in academia is adequate to prepare you for your career path?}
\end{itemize}

\end{document}